\documentclass[aps,pre,preprint,showpacs]{revtex4}
\usepackage{color}
\usepackage{colordvi}
\usepackage{epsfig}
\usepackage{bm,amssymb}
\usepackage{mathtools}

\begin{document}

\title{
Generating random networks that
consist of a single connected component
with a given degree distribution
}

\author{Ido Tishby}

\affiliation{Racah Institute of Physics, 
The Hebrew University, Jerusalem, 91904, Israel}

\author{Ofer Biham}

\affiliation{Racah Institute of Physics, 
The Hebrew University, Jerusalem, 91904, Israel}

\author{Eytan Katzav}

\affiliation{Racah Institute of Physics, 
The Hebrew University, Jerusalem, 91904, Israel}

\author{Reimer K\"uhn}

\affiliation{Mathematics Department, King's College London, 
Strand, London WC2R 2LS, United Kingdom}

\begin{abstract}

We present a method for the construction of 
ensembles of random networks 
that consist of a single connected component 
with a given degree distribution.
This approach extends the construction toolbox 
of random networks beyond the
configuration model framework, in which one 
controls the degree distribution but
not the number of components and their sizes.
Unlike configuration model networks, 
which are completely uncorrelated, the resulting
single-component networks exhibit 
degree-degree correlations.
Moreover, they are found to be disassortative, namely high-degree 
nodes tend to connect to low-degree nodes and vice versa.
We demonstrate the method for single-component networks
with ternary, exponential and power-law degree distributions.

\end{abstract}

\maketitle

\section{Introduction}

Network models provide a useful description of a
broad range of phenomena in the natural sciences and engineering
as well as in the economic and social sciences.
This realization has stimulated increasing interest 
in the structure of complex networks, 
and in the dynamical processes that take place on them 
\cite{Albert2002,Dorogovtsev2003,Dorogovtsev2008,Newman2010,Barrat2012,Hofstad2013,Havlin2010,Estrada2011,Latora2017}.
One of the central lines of inquiry 
has been concerned with the existence
of a  giant connected component
that is extensive in the network size.
In the case of 
Erd\H{o}s-R\'enyi (ER) networks,
the critical parameters for the 
emergence of a giant component in the thermodynamic limit  
were identified and the 
fraction of nodes that reside in the giant component was determined 
\cite{Erdos1959,Erdos1960,Erdos1961,Bollobas1984}. 
These studies were later extended to the broader class of
configuration model networks
\cite{Molloy1995,Molloy1998}. 
The configuration model framework enables one
to construct an ensemble of random networks whose 
degree sequences are drawn from a desired
degree distribution, with no degree-degree correlations. 
The resulting network ensemble 
is a maximum entropy ensemble
under the condition of the given degree distribution. 
A simple example of a configuration model network 
is the random regular graph,
in which all the nodes are of the same degree, $k=c$. 
For random regular graphs
with $c \ge 3$ the giant component encompasses the whole network
\cite{Bollobas2001}. 
However, in general, configuration model networks often
exhibit a coexistence between a
giant component, which is extensive in the network size, 
and many finite components,
which are non-extensive trees.
This can be exemplified by the case of ER networks,
which exhibit a Poisson degree distribution of the form

\begin{equation}
P(k) = \frac{e^{-c}c^k}{k!},
\label{eq:Poisson}
\end{equation}

\noindent
where 
$c=\langle K \rangle$ is the mean degree.
ER networks with $0 < c < 1$ consist of finite tree components.
At $c=1$ there is a percolation transition, above 
which the network exhibits a 
coexistence between the giant component 
and the finite components. 
In the asymptotic limit,
the size of the giant
component is $N_1 = g N$, where $N$ is the 
size of the whole network and the parameter $g=g(c)$,
which vanishes for $c \le 1$,
increases monotonically for $c > 1$. 
At $c = \ln N$ there is a 
second transition, above which the giant component
encompasses the entire network
\cite{Bollobas2001}. 
In the range of $1 < c < \ln N$, where the
giant and finite components coexist, the structure and statistical properties
of the giant component differ significantly from those of the whole network.
In particular, the degree distribution of the giant component differs from 
$P(k)$ and it exhibits degree-degree correlations.

Recently, we developed a theoretical framework 
for the analytical calculation of the
degree distribution and the degree-degree 
correlations in the giant component of
configuration model networks
\cite{Tishby2018}.
In particular, this framework provides an 
analytical expression for the
degree distribution of the giant component, 
denoted by $P(k|1)$, in terms of the
degree distribution $P(k)$ of the whole network.
We applied this approach to the most commonly 
studied configuration model networks, 
namely with Poisson, exponential and power-law degree distributions. 
We have shown that the degree distribution of the giant component 
enhances the weight of the high-degree nodes and depletes the low-degree nodes,
 with respect to the whole network. Moreover, we found that
the giant component is disassortative, namely high-degree nodes preferentially
connect to low-degree nodes and vice versa. This appears to be a crucial feature
that helps to maintain the integrity of the giant component.

In this paper we introduce a method
for the construction of ensembles of random networks that consist 
of a single connected component with a given degree distribution, $P(k|1)$.
This is done by inverting the equations that 
express the degree distribution of the
giant component $P(k|1)$ in terms of the 
degree distribution $P(k)$ of the whole network.
Constructing a configuration model network 
with the degree distribution $P(k)$ obtained
from the inversion process, its giant component is found to exhibit the 
desired degree distribution $P(k|1)$.
We apply this approach to the construction of 
ensembles of random networks that
consist of a single connected component with ternary, exponential and 
a power-law degree distributions.

The paper is organized as follows.
In Sec. II we present the configuration model network ensemble.
In Sec. III we present a method for the construction of a single-component network with
a given degree distribution.
In Sec. IV we analyze the properties of the resulting single-component networks.
In particular, we present analytical expressions for the degree-degree correlations and
the assortativity coefficient.
In Sec. V we apply this methodology for the construction of networks that
consist of a single connected component and exhibit ternary, exponential
and power-law distributions.
The results are discussed in Sec. VI and summarized in Sec. VII.

\section{The configuration model}

The configuration model network ensemble is an ensemble 
of uncorrelated random networks
whose degree sequences are drawn from
a given degree distribution, $P(k)$.
In theoretical studies one often considers the asymptotic case in which
the network size is infinite. 
In computer simulations, the network size $N$ is finite 
and the degree distribution is bounded from above and below
such that $k_{\rm min} \le k \le k_{\rm max}$.
For example, the commonly used choice of $k_{\rm min}=1$
eliminates the possibility of isolated nodes in the network. 
Choosing $k_{\rm min}=2$ also eliminates the leaf nodes.
Controlling the upper bound is important in the case of finite networks
with degree distributions that exhibit fat tails, 
such as power-law degree distributions.

The configuration model ensemble is a maximum entropy ensemble
under the condition that the degree distribution $P(k)$ is imposed
\cite{Newman2001,Newman2010}.
In this paper we focus on the case of undirected networks.
To generate a network instance drawn from an ensemble of
configuration model networks of $N$ nodes,
with a given degree distribution $P(k)$, one draws
the degrees of the $N$ nodes independently from 
$P(k)$.
This gives rise to a
degree sequence of the form
$k_1,k_2,\dots,k_N$
(where $\sum k_i$ must be even).
Configuration model networks do not exhibit degree-degree correlations,
which means that the conditional degree distribution 
of random neighbors of a random node of
degree $k$ satisfies
$P(k'|k)=k' P(k')/\langle K \rangle$
and does not depend on $k$. 
Also, the local structure of the network around 
a random node is typically a tree structure.
A central feature of configuration model networks and other random networks
above the percolation transition
is the small-world property, namely the fact that the mean distance scales like 
$\langle L \rangle \sim \ln N$.
Moreover, it was shown that scale-free networks 
for which 
$P(k) \propto k^{-\gamma}$ 
may be ultrasmall, 
depending on the exponent $\gamma$. 
In particular, for $2 < \gamma < 3$ their mean distance
scales like 
$\langle L \rangle \sim \ln \ln N$
\cite{Cohen2003}.

Configuration model networks in which $k_{\rm min}=1$
exhibit three different phases. In the sparse network limit,
below the percolation transition,
they consist of many finite tree components.
Above the percolation transition there is a coexistence of a
giant component and finite tree components.
In the dense network limit there is a second transition,
above which the giant component encompasses the whole network.
In this paper we focus on the intermediate domain in which the giant
and finite components coexist.
The size of the giant component is determined by the 
degree distribution $P(k)$.

\subsection{The construction of configuration model networks}

For the computer simulations presented below, we draw 
random network instances from an ensemble of configuration model
networks of $N$ nodes,
which follow a given degree distribution, $P(k)$.
For each network instance 
we generate a degree sequence of the form
$k_1, k_2,\dots,k_N$,
as described above.
For the discussion below it is convenient to list the degree
sequence in a decreasing order of the form
$k_1 \ge k_2 \ge \dots \ge k_N$.

It turns out that not every possible degree sequence is graphic,
namely admissible as a degree sequence of a network.
Therefore, before trying to construct a network with a given
degree sequence, one should first confirm
the graphicality of the degree sequence.
To be graphic, a degree sequence must satisfy two conditions.
The first condition is that the sum of the degrees is an even number,
namely
$\sum_{i=1}^N k_i = 2 L$,
where $L$ is an integer that represents
the number of edges in the network.
The second condition is expressed by the Erd{\H o}s-Gallai theorem,
which states that an ordered sequence of the form
$k_1 \ge k_2 \ge \dots \ge k_N$
is graphic if and only if the condition

\begin{equation}
\sum_{i=1}^n k_i \le n(n-1) + \sum_{i=n+1}^N \min (k_i,n)
\label{eq:EG}
\end{equation}

\noindent
holds for all values of $n$ in the range
$1 \le n \le N-1$
\cite{Erdos1960b,Choudum1986}.

A convenient way to construct a configuration model network 
is to prepare the $N$ nodes such that each node $i$ is 
connected to $k_i$ half edges or stubs
\cite{Newman2010}.
At each step of the construction, one connects a random pair of stubs that 
belong to two different nodes $i$ and $j$ 
that are not already connected,
forming an edge between them.
This procedure 
is repeated until all the stubs are exhausted.
The process may get stuck before completion in a case in which 
all the remaining stubs belong to the
same node or to pairs of nodes that are already connected.
In such case one needs to perform some random reconnections
in order to complete the construction.

\subsection{The degree distribution of the giant component}

Consider a configuration model network 
of $N$ nodes with a degree distribution, $P(k)$.
To obtain the probability $g$ that a random node
in the network belongs to the giant component,
one needs to first calculate the probability $\tilde g$,
that a random neighbor of a random node, $i$,
belongs to the giant component of the reduced network  
that does not include the node $i$.
The probability $\tilde g$ is determined by
\cite{Havlin2010}

\begin{equation}
1 - {\tilde g} = G_1(1 - {\tilde g}),
\label{eq:tg}
\end{equation}

\noindent
where

\begin{equation}
G_1(x) = \sum_{k=1}^{\infty}   x^{k-1}   {\widetilde P}(k) 
\label{eq:G1}
\end{equation}

\noindent
is the generating function of ${\widetilde P}(k)$, 
and

\begin{equation}
{\widetilde P}(k) = \frac{k}{\langle K \rangle} P(k)
\label{eq:tilde}
\end{equation}

\noindent
is the degree distribution of nodes that are sampled as 
random neighbors of random nodes.
Using $\tilde g$, one can then obtain the probability $g$ from the equation

\begin{equation}
g = 1 - G_0(1 - {\tilde g}),
\label{eq:g}
\end{equation}

\noindent
where

\begin{equation}
G_0(x) = \sum_{k=0}^{\infty}   x^{k}   P(k) 
\label{eq:G0}
\end{equation}

\noindent
is the generating function of $P(k)$.
Given that $G_0(x)$ and $G_1(x)$,
defined by
Eqs. (\ref{eq:G0})
and (\ref{eq:G1}),
respectively,
are probability generating functions, 
they satisfy $G_0(1)=G_1(1) =1$. 
This property entails that $\tilde g =0$ 
is always a solution of 
Eq  (\ref{eq:tg}). 
This (trivial) solution implies $g=0$ and describes a subcritical network, 
in which case the key question is, whether other solutions with 
$\tilde g  >0$, hence $g>0$, exist as well.

In configuration model networks that do not include any 
isolated nodes (of degree $k=0$) and leaf nodes (of degree $k=1$),
namely $k_{\rm min} \ge 2$,
the generating functions satisfy
$G_0(0) = 0$ 
and
$G_1(0)=0$.
This solution corresponds to the case where the giant
component encompasses the whole network and $g=\tilde g=1$.
This implies that in such networks both $x=0$ and $x=1$ are fixed points
of both $G_0(x)$ and $G_1(x)$. 
Furthermore, it can be shown that in networks whose degree distributions satisfy
the condition that 
$k_{\rm min} \ge 2$ 
and
$k_{\rm max} \ge 3$
there are no other (nontrivial) fixed points
for $G_0(x)$ and $G_1(x)$ with $0 < x < 1$
\cite{Bonneau2017}.
This means that in such networks the giant component encompasses the
whole network.
Here we are interested in configuration model networks that exhibit a 
coexistence between the giant and the finite components.
Such coexistence appears for degree distributions that support a non-trivial 
solution of Eq. (\ref{eq:tg}), in which $0 < \tilde g < 1$.
A necessary condition for such solution is the existence of leaf nodes
of degree $k=1$,
namely, $P(1) > 0$.
Therefore, we focus here on degree distributions in which $k_{\rm min}=1$.

For the analysis presented below
we introduce an indicator variable
$\Lambda \in \{0,1\}$,
where $\Lambda=1$ indicates that an event takes place on the
giant component and $\Lambda=0$ 
indicates that it happens on one of the finite components.
In this notation, the probability that a random node
resides on the giant component is
$P(\Lambda = 1) = g$,
and the probability that it resides on one of the finite components is
$P(\Lambda=0) = 1 - g$.
Similarly, the probabilities that a random neighbor of a random node resides
on the giant 
component is
$\widetilde P(\Lambda = 1) = \tilde g$
and the probability that it resides on one of the finite components is
$\widetilde P(\Lambda=0) = 1 - \tilde g$.

A node, $i$, of degree $k$ resides on the 
giant component if at least one of its $k$ neighbors
resides on the giant component of the reduced 
network from which $i$ is removed.
Therefore, the probability $g_k$ that a random node of degree $k$
resides on the giant component is given by

\begin{equation}
g_k = P(\Lambda = 1 | k)= 1 - (1 - \tilde g)^k,
\label{eq:L1k} 
\end{equation}

\noindent
while the probability that such node resides on one of the
finite components is

\begin{equation}
P(\Lambda = 0 | k) = 1 - g_k =  (1 - \tilde g)^k.
\label{eq:L0k}
\end{equation}

\noindent
Using Bayes' theorem, one can show that the degree distribution,
conditioned on the giant component, is given by
\cite{Tishby2018}

\begin{equation}
P(k | \Lambda =1) =
\frac{ 1- (1-\tilde g)^k}{g} P(k),
\label{eq:kL1}
\end{equation}

\noindent
while the degree distribution,
conditioned on the finite components,
is given by

\begin{equation}
P(k | \Lambda =0) =
\frac{ (1-\tilde g)^k}{1 - g} P(k).
\label{eq:kL0}
\end{equation}

\noindent
The mean degree of the giant component is 

\begin{equation}
\mathbb{E}[K | \Lambda=1] =
\frac{1-(1-\tilde g)^2}{g} 
\langle K \rangle,
\label{eq:EkL1}
\end{equation}

\noindent
while the mean degree on the finite components is

\begin{equation}
\mathbb{E}[K | \Lambda=0] =
\frac{(1-\tilde g)^2}{1-g} 
\langle K \rangle,
\label{eq:EkL0}
\end{equation}

\noindent
where

\begin{equation}
\langle K \rangle = \sum_{k=0}^{\infty} k P(k)
\end{equation}

\noindent
is the mean degree of the whole network.
In the rest of the paper, for the sake of brevity, 
we will drop the indicator $\Lambda$ and use $P(k|0)$ and $P(k|1)$ 
to denote the degree distribution on the finite components and on 
the giant component, respectively. Similarly, we will use 
$\mathbb{E}[K |0]$ ($\mathbb{E}[K |1]$) to denote the 
expected degree on the finite (giant) component.
It is interesting to mention that just above the percolation transition, 
when the giant component just emerges, $\mathbb{E}[K |1] \rightarrow 2$ 
\cite{Tishby2018,Tishby2018b}. 
This will be important in the rest of the paper, 
because it means that if one wants to generate a network that forms a single component 
with a given degree distribution $P(k|1)$, the
mean of this distribution must satisfy
$\mathbb{E}[K|1] \ge 2$.
From a different angle, a single tree component of $N$ nodes satisfies
$\mathbb{E}[K|1] = 2-2/N$
\cite{Katzav2018},
thus 
$\mathbb{E}[K|1] \rightarrow 2$
in the asymptotic limit.
Above the percolation transition cycles start to emerge in the giant
component, and 
$\mathbb{E}[K|1]$
gradually increases.
As the network becomes more dense, the fraction of nodes,
$g$, that reside on the giant component increases.
When $g \rightarrow 1$ the giant component encompasses the whole network.
The value of $\mathbb{E}[K|1]$ at which $g \rightarrow 1$ depends on the
degree distribution.

\subsection{The size of the giant component}

The expectation value of the size of the giant component of 
a configuration model of $N$ nodes with a degree distribution $P(k)$
is given by 

\begin{equation}
\langle N_1 \rangle =  N g,
\label{eq:N_1} 
\end{equation}

\noindent
where $g$ is given by Eq. (\ref{eq:g}).
However, in any single network instance the size $N_1$ of the giant
component may deviate from $\langle N_1 \rangle$.
Below we consider the distribution $P(N_1)$ 
of the sizes of the giant components obtained
in an ensemble of configuration model networks of $N$ nodes
with degree distribution $P(k)$.
To get a rough idea about the form of $P(N_1)$, one may assume,
for simplicity, that each node independently resides on the
giant component with probability $g$, with no correlations between
different nodes. In such case, $P(N_1)$ would follow a binomial
distribution that converges to a Gaussian distribution
whose mean is given by Eq. (\ref{eq:N_1}).
The variance of such a distribution is given by

\begin{equation}
{\rm Var}(N_1) = 
N \sum_{k=1}^{N-1} 
g_k (1-g_k) P(k),
\label{eq:VarN1}
\end{equation}

\noindent
where $g_k$ is given by Eq. (\ref{eq:L1k}).
In dense networks that exhibit a narrow degree distribution,
such that $g_k$ is only weakly dependent on $k$,
Eq. (\ref{eq:VarN1}) can be approximated by

\begin{equation}
{\rm Var}(N_1) = 
N g (1-g).
\label{eq:VarN1app}
\end{equation}

\noindent
In the case of ER networks 
\cite{Kang2016,Bollobas2013}, 
in which $P(k)$ is a Poisson distribution, 
as in Eq. (\ref{eq:Poisson}), 
it was shown that $P(N_1)$ is a Gaussian distribution
whose mean is given by
Eq. (\ref{eq:N_1})
and its variance is given by

\begin{equation}
{\rm Var}(N_1) = \frac{Ng(1-g)}{1 - \langle K \rangle(1-g)}.
\end{equation}

\noindent
For configuration model networks with other degree distributions there
are rigorous results for the size distribution of the giant component 
only in the weakly supercritical range 
\cite{Riordan2012,Bollobas2013},
which is just above the percolation phase transition. 
More precisely, in configuration model networks the percolation transition 
follows the Molloy-Reed criterion 
\cite{Molloy1995,Molloy1998}, 
namely, it takes place at  
$\langle K(K-1) \rangle/\langle K \rangle = 1$. 
Just above the transition, in the limit
$\epsilon=\langle K(K-1) \rangle/\langle K \rangle - 1 \rightarrow 0^{+}$,
the distribution
$P(N_1)$ is a Gaussian distribution
whose mean is 

\begin{equation}
\langle N_1 \rangle = \frac{2 \langle K \rangle^2}{\langle K(K-1)(K-2) \rangle} \epsilon N
\end{equation}

\noindent
and its variance is given by

\begin{equation}
{\rm Var}(N_1) = \frac{2 \langle K \rangle}{\epsilon} N.
\end{equation}

\noindent
This means that at the percolation transition the variance of $N_1$ diverges, 
and starts decreasing above the transition. There are no rigorous results in the 
full supercritical range, but following the ER case, it is plausible that the normality 
of $P(N_1)$ still holds, at least for a degree distribution $P(k)$ with a finite variance, 
while the variance ${\rm Var}(N_1)$ decreases.
The main conclusion of this discussion is that  
sufficiently far above the percolation transition, where the giant component is 
not too small, the size fluctuations of the giant component 
become negligible as $N$ is increased.

\section{The construction of a single-component network with a given degree distribution}

Here we present a method for the construction of a network that consists of a single component
whose degree sequence is effectively drawn from
a given degree distribution, denoted by $P(k|1)$. 
The approach is based on the construction of a  configuration model network  
whose degree sequence is drawn from a suitable   
degree distribution $P(k)$, such that its giant component 
exhibits the desired degree distribution, $P(k|1)$. 

Inverting Eq. (\ref{eq:kL1}) we find that in
order to obtain a giant component whose degree distribution is
$P(k|1)$, the degree distribution of the whole network should be

\begin{equation}
P(k) = \frac{g}{1-(1-\tilde g)^k} P(k|1),
\label{eq:pk2}
\end{equation}

\noindent
where $\tilde g$ is given by Eq. (\ref{eq:tg}) and
$g$ is given by Eq. (\ref{eq:g}).
The mean degree of the whole network will thus be

\begin{equation}
\langle K \rangle = 
\sum_{k=1}^{\infty}
\frac{g k }{1-(1-\tilde g)^k} P(k|1).
\end{equation}

\noindent
In order to obtain an ensemble of single-component networks
whose mean size is $\langle N_1 \rangle$, the size of the configuration 
model networks from which these giant components are obtained should be

\begin{equation}
N  = \frac{\langle N_1 \rangle}{g}.
\label{eq:N1g}
\end{equation}

\noindent
For the analysis below it is useful to introduce the
generating functions for the degree distribution
conditioned on the giant component, namely

\begin{equation}
G_0^1(x) =  \sum_{k=1}^{\infty}
x^{k} P(k|1)
\label{eq:G01}
\end{equation}

\noindent
and

\begin{equation}
G_1^1(x) =  \sum_{k=1}^{\infty} 
\frac{ k x^{k-1}  }{ {\mathbb E}[K|1]} P(k|1).
\label{eq:G11}
\end{equation}

\noindent
These generating functions are related to each other by the equation

\begin{equation}
G_1^1(x) = \frac{ \frac{d  }{dx} G_0^1(x)  }{  \frac{d }{dx} G_0^1(x)  \vert_{x=1}} .
\label{eq:G0G1}
\end{equation}

\noindent
In order to calculate the probability $\tilde g$, we utilize Eq. (\ref{eq:tg}),
where we express $P(k)$ and $\langle K \rangle$
in terms of $P(k|1)$, and obtain

\begin{equation}
1 - \tilde g = \frac{ \sum\limits_{k=1}^{\infty} \frac{ k\left(1-\tilde g \right)^{k-1} }
{1 - (1-\tilde g)^k} P(k|1) }
{ \sum\limits_{k=1}^{\infty} \frac{ k }{1 - (1-\tilde g)^k} P(k|1) }.
\end{equation}

\noindent
Using the Taylor expansion of $(1-x)^{-1}$, which takes the form

\begin{equation}
\frac{1}{1-x} = \sum_{n=0}^{\infty} x^n,
\label{eq:taylor}
\end{equation}

\noindent
where $0 < x < 1$, to express the term $1/[1-(1-\tilde g)^k]$
as a power series in $(1-\tilde g)^k$, we obtain

\begin{equation}
1 - \tilde g = 
\frac{ \sum\limits_{k=1}^{\infty}  k (1-\tilde g)^{k-1} \sum\limits_{n=0}^{\infty} (1-\tilde g)^{kn} P(k|1) }
{ \sum\limits_{k=1}^{\infty} k \sum\limits_{n=0}^{\infty} (1-\tilde g)^{kn} P(k|1) }.
\end{equation}

\noindent
Multiplying both sides by $1-\tilde g$ and exchanging the order of summations in
the numerator and denominator, we obtain

\begin{equation}
(1 - \tilde g)^2 = 
\frac{ 
\sum\limits_{n=1}^{\infty} (1-\tilde g)^n 
\sum\limits_{k=1}^{\infty} k (1-\tilde g)^{n(k-1)} P(k|1) }
{ \sum\limits_{n=0}^{\infty} (1-\tilde g)^n \sum\limits_{k=1}^{\infty} k  (1-\tilde g)^{n(k-1)} P(k|1) }.
\end{equation}

\noindent
Adding and subtracting the $n=0$ term in the numerator, 
this equation can be expressed in the form

\begin{equation}
(1 - \tilde g)^2 = 
1 -
\frac{ {\mathbb E}[K|1] }
{ \sum\limits_{n=0}^{\infty} (1-\tilde g)^n \sum\limits_{k=1}^{\infty} k  (1-\tilde g)^{n(k-1)} P(k|1) }.
\label{eq:1mtgs}
\end{equation}

\noindent
Using the generating function $G_1^1(x)$,
Eq. (\ref{eq:1mtgs}) can be written in the form

\begin{equation}
(1 - \tilde g)^2 = 
1 -
\frac{ 1 }
{ \sum\limits_{n=0}^{\infty} (1-\tilde g)^n G_1^1[(1-\tilde g)^n] },
\end{equation}

\noindent
or in the form

\begin{equation}
 \tilde g (2-\tilde g) \sum_{n=0}^{\infty} 
(1-\tilde g)^n G_1^1[(1-\tilde g)^n] = 1.
\label{eq:g2mg}
\end{equation}

\noindent
This is an implicit equation that should be solved in order to obtain the 
parameter $\tilde g$. For some degree distributions one can obtain a closed
form analytical expression for $\tilde g$, while for other distributions it should
be calculated numerically. A useful approximation scheme would
be to replace the sum in Eq. (\ref{eq:g2mg}) by an integral. 
To improve the accuracy of this approximation, it is useful to first
separate the $n=0$ and the $n=1$ terms from the rest of the
sum and obtain

\begin{equation}
 \tilde g (2-\tilde g) 
\left[ 1 + (1-\tilde g) G_1^1(1-\tilde g) +
\sum_{n=2}^{\infty} 
(1-\tilde g)^n G_1^1[(1-\tilde g)^n] 
\right]
= 1.
\label{eq:g2mg12}
\end{equation}

\noindent
Using Eq. (\ref{eq:G0G1}) we find that

\begin{equation}
x^n G_1^1(x^n) =  \frac{ 
\frac{\partial}{\partial n}
[G_0^1(x^n)] }{  {\mathbb E}[K|1] \ln x }.
\label{eq:xnGxn}
\end{equation}

\noindent
Replacing the sum $\sum_{n=2}^{\infty}$ in 
Eq. (\ref{eq:g2mg12})
by an integral of the form $\int_{3/2}^{\infty} dn$
and carrying out the integration using Eq. (\ref{eq:xnGxn}),
we obtain

\begin{equation}
\tilde g (2-\tilde g)
\left[ 1 + (1-\tilde g) G_1^1(1-\tilde g)
- \frac{G_0^1\left[ (1-\tilde g)^{3/2} \right]}{{\mathbb E}[K|1] \ln (1-\tilde g)} \right]
=1.
\label{eq:gtimplicit}
\end{equation}

\noindent
This equation is easier to handle than Eq. (\ref{eq:g2mg}), 
although usually it can be solved only numerically. 
Other, more precise schemes, could be devised by treating 
more individual terms of the sum in Eq. (\ref{eq:g2mg}) separately, 
say up to $n=2$ or $n=3$, and approximating the tail of the sum by an integral. 
Our experience tells us that for the cases considered in this paper 
using the $n=1$ scheme provides values of $\tilde g$ that differ 
by at most a few percents from the exact value.

Once the parameter $\tilde g$ is known, the parameter $g$ can be obtained from

\begin{equation}
1 - g = \sum_{k=0}^{\infty}
\frac{g  (1-\tilde g)^k }{1 - (1-\tilde g)^k} P(k|1).
\end{equation}

\noindent
Extracting $g$ we obtain

\begin{equation}
g =  \frac{1}{ 1 + \sum\limits_{k=0}^{\infty} \frac{(1-\tilde g)^k}{1 - (1-\tilde g)^k} P(k|1) }.
\end{equation}

\noindent
Expanding the denominator according to Eq. (\ref{eq:taylor})
and exchanging the order of the summations, we obtain

\begin{equation}
g = \frac{1}{ 1 + \sum\limits_{n=1}^{\infty} G_0^1[(1-\tilde g)^n] }.
\label{eq:gshort}
\end{equation}

\noindent
To conclude, in order to obtain an ensemble of single-component networks 
whose mean size is $\langle N_1 \rangle$, with 
degree sequences that are effectively drawn from $P(k|1)$, one constructs
an ensemble of configuration model networks whose size $N$ is given by Eq. (\ref{eq:N1g})
and its degree distribution $P(k)$ is given by Eq. (\ref{eq:pk2}). 
The giant components of these networks are the desired
single component networks. 
The mean degree $\langle K \rangle$ of the configuration model networks is

\begin{equation}
\langle K \rangle = 
\frac{g}{1-(1-\tilde g)^2}
\mathbb{E}[K | 1].
\label{eq:<K>_EK}
\end{equation}

Note that it is also possible to control the exact size of the single-component network.
Consider the case in which the desired size of a given instance of the single-component
network is $\lfloor \langle N_1 \rangle \rfloor$, namely the integer part
of $\langle N_1 \rangle$. In a case in which the size of the giant component $n_1$ came out 
smaller than $\lfloor \langle N_1 \rangle \rfloor$, one should add nodes to the configuration
model network until the giant component will reach the desired size. The degrees of the
added nodes are drawn from $P(k)$. To add a node of an even degree $k$ to the network
one picks randomly $k/2$ edges that connect $k$ distinct nodes. 
One then cuts each edge in the middle to generate $k$ stubs. The
$k$ stubs of the new node are then connected to these $k$ stubs. In the case of nodes
of odd degrees, $k$ and $k'$, one picks randomly $(k+k')/2$ edges and cuts them in the middle
to generate $k+k'$ stubs. The stubs of the two new nodes are then connected randomly to 
these $k+k'$ stubs. In a case in which $n_1$ came out larger than $\lfloor \langle N_1 \rangle \rfloor$
one should delete random nodes (one at a time for even-degree nodes and in pairs
for odd-degree nodes), until the giant component is reduced to the desired size. 
The open stubs that remain from the edges of each deleted node are
then randomly connected to each other in pairs. 

\section{Properties of single component random networks}

Unlike configuration model networks that are completely uncorrelated, 
their giant components exhibit degree-degree correlations. 
In particular, following the observation made in Ref. \cite{Tishby2018}
that the giant components are disassortative, below we prove this property.
Interestingly, this observation has been recently demonstrated in
percolating clusters 
\cite{Mizutaka2018}.

The joint degree distribution of a pair of adjacent nodes
in a configuration model network with degree distribution $P(k)$ is given by
\cite{Tishby2018}

\begin{equation}
\widehat P(k,k'|1) = \frac{1 - (1-\tilde g)^{k+k'-2}}{1-(1-\tilde g)^2} 
\frac{k}{\langle K \rangle} P(k) \frac{k'}{\langle K \rangle} P(k').
\end{equation}

\noindent
Expressing $P(k)$ and $P(k')$ in terms of $P(k|1)$ and $P(k'|1)$, respectively,
using Eq. (\ref{eq:pk2}), we obtain

\begin{equation}
\widehat P(k,k'|1) =
W(k,k')
\frac{k}{{\mathbb E}[k|1]} P(k|1)
\frac{k'}{{\mathbb E}[k|1]} P(k'|1),
\end{equation}

\noindent
where

\begin{equation}
W(k,k') =
\tilde g (2-\tilde g) 
\frac{1-(1-\tilde g)^{k+k'-2}}{[1-(1-\tilde g)^k][1-(1-\tilde g)^{k'}]}
\end{equation}

\noindent
accounts for the degree-degree correlations between adjacent nodes.
For example, $W(1,1)=0$, reflecting the fact that pairs of nodes of degree $k=1$
on the giant component cannot share an edge, because in such case they will
form an isolated dimer. Also, one can verify that $W(k,2)=1$ for all values of
$k \ge 1$. This means that nodes of degree $k=2$ are distributed
randomly in the giant component and are not correlated to the degrees
of their neighboring nodes. The degree-degree correlations between nodes
of degree $k \ge 3$ and leaf nodes of degree $k'=1$ is given by

\begin{equation}
W(k,1) = 1 + \frac{1 - \tilde g - (1-\tilde g)^{k-1}}{1-(1-\tilde g)^k} > 1.
\end{equation}

\noindent
Thus, there is a positive correlation between leaf nodes and nodes of degree
$k \ge 3$. Moreover, the correlation becomes stronger as $k$ increases.

Below we show that 
$W(k,k') \le 1$ 
for
for $k,k' \ge 3$,
hence the degree-degree correlations between pairs of nodes
of degrees $k,k' \ge 3$ are negative.
To this end we denote $\tilde h = 1 - \tilde g$, which satisfies
$0 < \tilde h < 1$. 
Expressing $W(k,k')$ in terms of $\tilde h$, we obtain

\begin{equation}
W(k,k';\tilde h) =
(1-{\tilde h}^2)
\frac{1-{\tilde h}^{k+k'-2}}{(1-{\tilde h}^k)(1-{\tilde h}^{k'})}.
\end{equation}

\noindent
The diagonal terms, obtained for $k=k'$, are given by

\begin{equation}
f(k;\tilde h) = W(k,k;\tilde h) =
(1-{\tilde h}^2)
\frac{1-{\tilde h}^{2k-2}}{(1-{\tilde h}^k)^2}.
\end{equation}

\noindent
For $k=3$ we obtain

\begin{equation}
f(k=3;\tilde h) = \frac{ (1+\tilde h)^2 (1+\tilde h^2) }{( 1+\tilde h + \tilde h^2 )^2}.
\end{equation}

\noindent
Differentiating $f(k=3;\tilde h)$ with respect to $\tilde h$, we obtain

\begin{equation}
\frac{\partial}{\partial \tilde h} f(k=3;\tilde h) =
- \frac{2 \tilde h (1-\tilde h^2) }{(1+\tilde h+\tilde h^2)^3}
< 0,
\end{equation}

\noindent
for $0 < \tilde h < 1$.
Therefore, the function $f(k=3;\tilde h)$ is a monotonically decreasing 
function of $\tilde h$.
This implies that

\begin{equation}
f(k=3;\tilde h) \le f(k=3;\tilde h=0) =1,
\end{equation}

\noindent
with equality taking place only at $\tilde h=0$.
Considering the degree, $k$, as a continuous variable and taking
the derivative of $f(k;\tilde h)$ with respect to $k$,
we obtain

\begin{equation}
\frac{\partial}{\partial k} f(k;\tilde h) =
- \frac{ 2 \tilde h^k (1-\tilde h^2)(1-\tilde h^{k-2}) \ln \left( \frac{1}{\tilde h} \right) }
{(1-\tilde h^k)^3} < 0
\end{equation}

\noindent
for $k > 2$ and $0 < \tilde h < 1$.
This means that $f(k;\tilde h)$ is a monotonically
decreasing function in both $k$ and $\tilde h$.
We thus conclude that $W(k,k) < 1$ for all values of $k \ge 3$
and $0 < \tilde h < 1$.
In order to show that $W(k,k') < 1$ for all $k,k' \ge 3$,
it is sufficient to show that under these conditions $W(k,k')$ is 
a monotonically decreasing function of $k'$ for all values of
$0 < \tilde h < 1$.
This is shown by differentiating $W(k,k';\tilde h)$ with respect to $k'$,
which leads to

\begin{equation}
\frac{ \partial }{ \partial k'}
W(k,k';\tilde h) = - \frac{ \tilde h^{k'} (1-\tilde h^2)(1-\tilde h^{k-2}) \ln \left( \frac{1}{\tilde h} \right) }
{ (1-\tilde h^k)(1-\tilde h^{k'})^2 } < 0
\end{equation}

\noindent
where $k > 2$ and $0 < \tilde h < 1$.
This means that for any combination of $k,k' \ge 3$, where $k' > k$,
the correlation function $W(k,k')$ satisfies
$W(k,k') < W(k,k) < 1$.
We thus conclude that pairs of adjacent nodes of degrees $k,k' \ge 3$
exhibit negative degree-degree correlations.

The probability that a node connected to a random edge 
in the giant component is of degree $k$ is given by
\cite{Tishby2018}

\begin{equation}
\widehat P(k|1) = \frac{k}{ {\mathbb E}[K|1] } P(k|1).
\end{equation}

\noindent
The assortativity coefficient 
\cite{Newman2002b}
of the giant component is given by
\cite{Tishby2018}

\begin{equation}
r = \frac{ \sum_{k,k' \ge 2} (k-1)(k'-1) \widehat P(k,k'|1) 
- \left[ \sum_{k \ge 2} (k-1) \widehat P(k|1) \right]^2 }
{ \sum_{k \ge 2} (k-1)^2 \widehat P(k|1) 
- \left[ \sum_{k \ge 2} (k-1) \widehat P(k|1) \right]^2 }.
\end{equation}

\noindent
Since the degree-degree correlations  between pairs of adjacent nodes
of degrees $k,k' \ge 3$ are negative, the assortativity coefficient of the giant
component must satisfy $r < 0$. This is an essential property of the
giant components of configuration model networks, which is required
in order to maintain the integrity of the giant component.

\section{Applications to specific network models}

In this section we apply the methodology developed above for the 
construction of networks that consist of a single connected component, with a 
prescribed degree distribution, $P(k|1)$, for some popular ensembles 
of random networks.

\subsection{Construction of a single-component network with a ternary degree distribution}

The properties of the giant component  of a random network are  
sensitive to the abundance of nodes of low degrees, particularly
nodes of degree $k=1$ (leaf nodes) and $k=2$.
Nodes of degree $k=0$ (isolated nodes) are excluded from the
giant component and their weight in the degree distribution
of the whole network has no effect on the properties of the giant component.
Therefore, it is useful to consider a simple configuration model
in which all nodes are restricted to a small number
of low degrees. 
Here we consider a configuration model network with a ternary 
degree distribution of the form
\cite{Newman2010}

\begin{equation}
P(k) = p_1 \delta_{k,1} + p_2 \delta_{k,2} + p_3 \delta_{k,3},
\label{eq:ternary}
\end{equation}

\noindent
where $\delta_{k,n}$ is the Kronecker delta,
and 
$p_1+p_2+p_3=1$.
The mean degree of such network is given by

\begin{equation}
\langle K \rangle = p_1 + 2 p_2 + 3 p_3.
\end{equation}

\noindent
The generating functions of the degree distribution are

\begin{equation}
G_0(x) = p_1 x + p_2 x^2 + p_3 x^3,
\label{eq:G0tri}
\end{equation}

\noindent
and

\begin{equation}
G_1(x) = \frac{ p_1  + 2 p_2 x + 3 p_3 x^2}{p_1 + 2 p_2 + 3 p_3}.
\label{eq:G1tri}
\end{equation}

\noindent
Solving Eq. (\ref{eq:tg}) for $\tilde g$, with $G_1(x)$ given by 
Eq. (\ref{eq:G1tri}), 
we find that

\begin{equation}
\tilde g = 
\begin{dcases}
0 &  \ \ \ \  p_3 \le \frac{p_1}{3} \\
1 - \frac{p_1}{3p_3} &  \ \ \ \  p_3 > \frac{p_1}{3}.
\end{dcases}
\end{equation}

\noindent
Using Eq. (\ref{eq:g}) to evaluate the parameter $g$, 
where $G_0(x)$ is given by 
Eq. (\ref{eq:G0tri}),
we find that

\begin{equation}
g = 
\begin{dcases}
0 & \ \ \ \   p_3 \le \frac{p_1}{3} \\
1 
- \left( \frac{p_1}{3p_3} \right) p_1
- \left( \frac{p_1}{3 p_3} \right)^2  p_2
- \left( \frac{p_1}{3 p_3} \right)^3 p_3 &
\ \ \ \   p_3 > \frac{p_1}{3}.
\end{dcases}
\end{equation}

\noindent
Thus, the percolation threshold is located at $p_3 = p_1/3$.
This can be understood intuitively by recalling that the finite 
components exhibit a tree structure. In a tree that
includes a single node of degree $k=3$ there must be
three leaf nodes of degree $k=1$. 
In the giant component, 
which includes cycles,
there must be more than one node of
degree $3$ for every three nodes of degree $1$.
This is not likely to occur in a case in which $p_3 < p_1/3$.
Using the normalization condition, we find that for any given
value of $p_2$, a giant component exists for 
$p_3 > (1-p_2)/4$.

Using Eq. (\ref{eq:kL1}), we obtain the
degree distribution of the giant component,
which is given by

\begin{equation}
P(k | 1) = 
\left[
\frac{ 1 - \left( \frac{p_1}{3 p_3} \right)^k }
{ 1 - \left( \frac{p_1 }{3 p_3} \right) p_1
- \left( \frac{p_1 }{3 p_3 } \right)^2 p_2
- \left( \frac{p_1 }{3 p_3} \right)^3  p_3 }
\right] 
P(k),
\label{eq:Pk1ter}
\end{equation}

\noindent
where $k=1, 2, 3$ and $P(k)$ is given by Eq. (\ref{eq:ternary}).

These results enable us to construct a giant connected component with 
a desired ternary degree distribution, given by
$P(k|1)$, $k=1,2,3$,
where
$\sum_{k=1}^3 P(k|1)=1$.
To this aim, we need to express the degree distribution $P(k)$ of the whole network,
given by Eq. (\ref{eq:ternary}),
in terms of the given degree distribution $P(k|1)$ of the giant component.
We should first evaluate the parameter $\tilde g$,
which is given by

\begin{equation}
\tilde g = 1 - \frac{p_1}{3 p_3}.
\end{equation}

\noindent
Using Eq. (\ref{eq:Pk1ter}) 
to calculate the ratio $P(1|1)/P(3|1)$,
we obtain

\begin{equation}
\frac{P(1|1)}{3 P(3|1)} = 
\frac{1}{1 + \left( \frac{p_1}{3 p_3} \right) + \left( \frac{p_1}{3 p_3} \right)^2 }
\ \frac{p_1}{3 p_3} 
\end{equation}

\noindent
Solving for $p_1/(3 p_3)$ we obtain

\begin{equation}
\frac{p_1}{3 p_3} = \frac{1}{2}
\left[
\frac{3 P(3|1)}{ P(1|1)}  - 1
- 
\sqrt{
\left( \frac{3 P(3|1)}{P(1|1)}  + 1 \right)
 \left( \frac{3 P(3|1)}{P(1|1)} - 3 \right)
}
\right].
\end{equation}

\noindent
Therefore

\begin{equation}
\tilde g = \frac{1}{2}
\left[
3 -
\frac{3 P(3|1)}{ P(1|1)}  
+
\sqrt{
\left( \frac{3 P(3|1)}{P(1|1)}  + 1 \right)
 \left( \frac{3 P(3|1)}{P(1|1)} - 3 \right)
}
\right].
\label{eq:tgternary3}
\end{equation}

\noindent
The next step is to evaluate the parameter $g$,
which is given by

\begin{equation}
g = \frac{1}{ 1 + \sum\limits_{k=1}^{3} \frac{(1-\tilde g)^k}{1 - (1-\tilde g)^k} P(k|1) }.
\end{equation}

\noindent
Simplifying the expression we obtain

\begin{equation}
g = \frac{\tilde g}{P(1|1) + \frac{1}{2-\tilde g}P(2|1) + \frac{1}{3-3 \tilde g+\tilde g^2}P(3|1)}.
\label{eq:gternary}
\end{equation}

\noindent
Using the normalization condition of the probabilities $P(k|1)$ to express $P(2|1)$ in
terms of $P(1|1)$ and $P(3|1)$ we obtain

\begin{equation}
g = \frac{\tilde g (2-\tilde g)}{1+(1-\tilde g)P(1|1) - \frac{(1-\tilde g)^2}{3-3 \tilde g+\tilde g^2}P(3|1)}.
\label{eq:gternary3}
\end{equation}

\noindent
The degree distribution of the whole network is given by
Eq. (\ref{eq:ternary}), where

\begin{eqnarray}
p_1 &=& \frac{g}{\tilde g} P(1|1)
\nonumber \\
p_2 &=& \frac{g}{\tilde g (2-\tilde g)} P(2|1)
\nonumber \\
p_3 &=& \frac{g}{\tilde g (3 - 3 \tilde g + \tilde g^2)} P(3|1).
\label{eq:p1p2p3}
\end{eqnarray}

\noindent
Thus, in order to obtain an ensemble of single-component networks 
of mean size $\langle N_1 \rangle$,
whose degree sequences are drawn from a given ternary degree distribution $P(k|1)$,
one generates an ensemble of configuration model networks with a degree distribution $P(k)$,
given by Eq. (\ref{eq:ternary}), where $p_1,p_2$ and $p_3$ are given by 
Eq. (\ref{eq:p1p2p3}). The size of the configuration model networks should be
$N=\langle N_1 \rangle/g$, where $g$ is given by Eq. (\ref{eq:gternary}).

\begin{figure}
\begin{center}
\includegraphics[width=7cm]{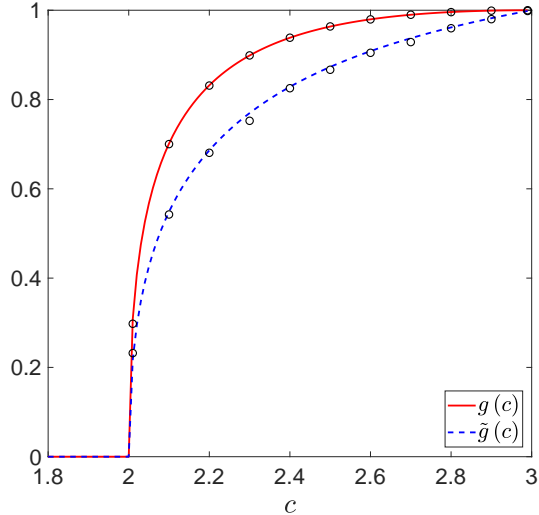} 
\end{center}
\caption{
(Color online)
Analytical results for the fraction of nodes $g$ (solid line), 
and the fraction of random neighbors of random nodes, 
$\tilde g$ (dashed line),
that reside on the giant component, in a configuration model network
whose giant component exhibits a ternary degree distribution
$P(k|1)$, expressed by Eq. (\ref{eq:Pk1ter}), with $P(K=2|1)=0$,
as a function of the mean degree $c={\mathbb E}[K|1]$
of the giant component.
The simulation results (circles),
obtained for $N=10^4$,
are in very good agreement with
the analytical results.
}
\label{fig:1}
\end{figure}

In Fig. \ref{fig:1} we present analytical results for 
the probability $g$, 
obtained from Eq. (\ref{eq:gternary3}),
that a randomly selected node
resides on the giant component (solid line),
in a configuration model network whose giant component
exhibits a ternary degree distribution with $P(K=2|1)=0$,
as a function of the mean degree $c={\mathbb E}[K|1]$
of the giant component.
We also show the probability $\tilde g$,
obtained from Eq. (\ref{eq:tgternary3}),
that a random neighbor of 
a random node resides on the giant component (dashed line).
As discussed above, both $g$ and $\tilde g$ vanish for $c<2$, 
since there are no giant components with mean degrees
smaller than $2$. 
For $c>2$ both $g$ and $\tilde g$ exhibit a steep rise
as $c$ is increased, reaching $g=\tilde g=1$ at $c=3$,
where the giant component encompasses the whole network.
The results obtained from computer simulations (circles) 
with $N=10^4$ are found
to be in very good agreement with the analytical results.

\subsection{Construction of a single component network with an exponential degree distribution}

Consider a configuration model network 
whose giant component exhibits an exponential degree
distribution of the form

\begin{equation}
P(k|1) = A e^{- \alpha k},
\end{equation}

\noindent
where $k \ge k_{\rm min}$. 
Here we focus on the case of 
$k_{\rm min}=1$, 
for which the normalization 
factor is $A=e^{\alpha} - 1$.
The mean degree is given by

\begin{equation}
c={\mathbb E}[K|1] = \frac{1}{1 - e^{- \alpha}}.
\end{equation}

\noindent
For the analysis below, it is convenient to parametrize the
degree distribution in terms of the mean degree $c$.
Plugging in 
$\alpha = \ln c - \ln (c-1)$ 
we obtain

\begin{equation}
P(k|1) = \frac{1}{c} \left( \frac{c-1}{c} \right)^{k-1},
\label{eq:exp}
\end{equation}

\noindent
where $k \ge 1$.
The mean degree of nodes that reside on the giant component is
${\mathbb E}[K|1]=c$.
As noted above, a giant component exists only for $c \ge 2$.
This implies that $\alpha$ must satisfy the condition
$\alpha \le \ln 2$.
Inserting $P(k|1)$ from Eq. (\ref{eq:exp}) into Eqs. 
(\ref{eq:G01}) and (\ref{eq:G11})
and carrying out the summations, we find that the
generating functions for a giant component with an exponential
degree distribution take the form

\begin{equation}
G_0^1(x) = \frac{x}{c - x(c-1)}
\label{eq:G01exp}
\end{equation}

\noindent
and

\begin{equation}
G_1^1(x) = \frac{1}{\left[ c + (1-c) x \right]^2}.
\label{eq:G11exp}
\end{equation}

\noindent
Plugging in $x=(1-\tilde g)^n$ in Eq. (\ref{eq:G11exp}) and inserting
the result into Eq. (\ref{eq:g2mg}), we obtain that
$\tilde g$ is given by

\begin{equation}
\tilde g (2-\tilde g) \sum_{n=0}^{\infty} \frac{ (1-\tilde g)^n }
{ \left[ c + (1-c) (1-\tilde g)^n \right]^2 } = 1.
\end{equation}

\noindent
This is an implicit equation for $\tilde g$ in terms of the mean degree $c$, 
which is essentially equivalent to Eq. (\ref{eq:g2mg}) for the case 
of the exponential distribution.
It should be solved numerically
in order to obtain $\tilde g = \tilde g(c)$.
Following the general approximation scheme presented in section VI we 
solve instead Eq. (\ref{eq:gtimplicit}), which for the exponential distribution 
case can be written explicitly in the following simpler form

\begin{equation}
\tilde g (2-\tilde g) \left\{ 
1 + \frac{1-\tilde g}{ (1 - \tilde g + c \tilde g)^2}
+ 
\frac{(1-\tilde g)^{3/2}}{ c \left[ c + (1-c) (1-\tilde g)^{3/2} \right] }
\right\}
= 1.
\label{eq:gtexp}
\end{equation}

\noindent

To calculate the parameter $g$, we use Eq. (\ref{eq:gshort}).
Plugging in the generating function $G_0^1(x)$ 
of the exponential degree distribution,
given by Eq. (\ref{eq:G01exp}), we obtain

\begin{equation}
g =  \left[ 1 + \sum\limits_{n=1}^{\infty}
\frac{(1-\tilde g)^n}{c-(c-1)(1-\tilde g)^n} \right]^{-1},
\label{eq:gexp1}
\end{equation}

\noindent
where $\tilde g$ is given by Eq. (\ref{eq:gtexp}).
In the case of the exponential distribution we have a useful 
approximation scheme which is similar to the one used in the 
self-consistent equation for $\tilde g$. This amounts to 
separating the first term from the rest of the sum in Eq. (\ref{eq:gexp1}), and replacing the
sum by an integral. This yields

\begin{equation}
g =  \left[ 1 + \frac{1-\tilde g}{c-(c-1)(1-\tilde g)} 
+ \int_{3/2}^{\infty}
\frac{(1-\tilde g)^n}{c-(c-1)(1-\tilde g)^n} dn \right]^{-1},
\label{eq:gexp2}
\end{equation}

\noindent
Carrying out the integration, we obtain

\begin{equation}
g =  \left[ 1 + \frac{1-\tilde g}{c-(c-1)(1-\tilde g)} 
+ \frac{ \ln \left[{1-\left(\frac{c-1}{c}\right)(1-\tilde g)^{3/2}}\right]}{(c-1) \ln (1-\tilde g)}  \right]^{-1}.
\label{eq:gexp}
\end{equation}

\noindent
It turns out that this expression is precise within less than one percent compared to
the full expression (\ref{eq:gexp1}), even next to the percolation transition.

In order to obtain a single component network of $N_1$ nodes with a given exponential degree 
distribution, $P(k|1)$, one generates a configuration model network with the degree
distribution $P(k)$, 
given by Eq. (\ref{eq:pk2}),
where $\tilde g$ is given by Eq. (\ref{eq:gtexp}),
$g$ is given by Eq. (\ref{eq:gexp})
and $P(k|1)$ is given by Eq. (\ref{eq:exp}).

\begin{figure}
\begin{center}
\includegraphics[width=8cm]{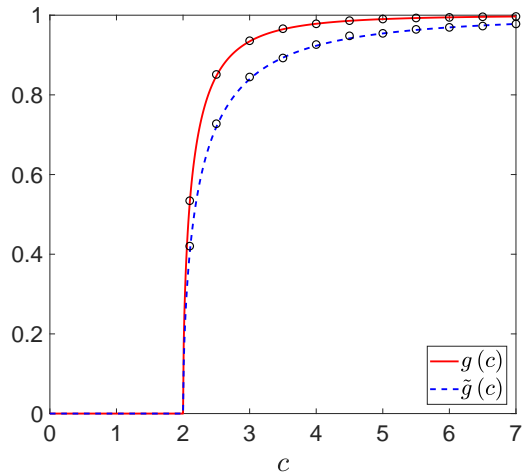} 
\end{center}
\caption{
(Color online)
The fraction of nodes, $g$ (solid line), 
and the fraction of random neighbors of random nodes $\tilde g$ (dashed line),
that reside on the giant component, in a configuration model network
whose giant component exhibits an exponential degree distribution,
$P(k|1)$, expressed by Eq. (\ref{eq:exp}),
as a function of the mean degree $c={\mathbb E}[K|1]$
of the giant component.
As discussed in the text, the minimal value of the mean degree of the 
giant component 
is $c=2$.
Thus, for $c<2$ both $g=0$ and $\tilde g=0$, while for $c>2$ 
the parameters $g$ and $\tilde g$ quickly increase.
The simulation results (circles),
obtained for $N=10^4$,
are in very good agreement with
the analytical results.
}
\label{fig:2}
\end{figure}

In Fig. \ref{fig:2} we present analytical results for 
the probability $g$, 
obtained from Eq. (\ref{eq:gexp}),
that a randomly selected node
resides on the giant component (solid line),
in a configuration model network whose giant component
exhibits an exponential degree distribution,
as a function of the mean degree $c={\mathbb E}[K|1]$
of the giant component.
We also show analytical results for the probability, $\tilde g$, 
obtained from Eq. (\ref{eq:gtexp}),
that a random neighbor of 
a random node resides on the giant component (dashed line).
As in the case of the ternary degree distribution, both $g$ and $\tilde g$ vanish for $c<2$, 
while for $c>2$ 
they exhibit a steep rise
as $c$ is increased.
The results of computer simulations (circles) 
with $N=10^4$
are in very good agreement with the analytical results. 

\begin{figure}
\begin{center}
\includegraphics[width=6.4cm]{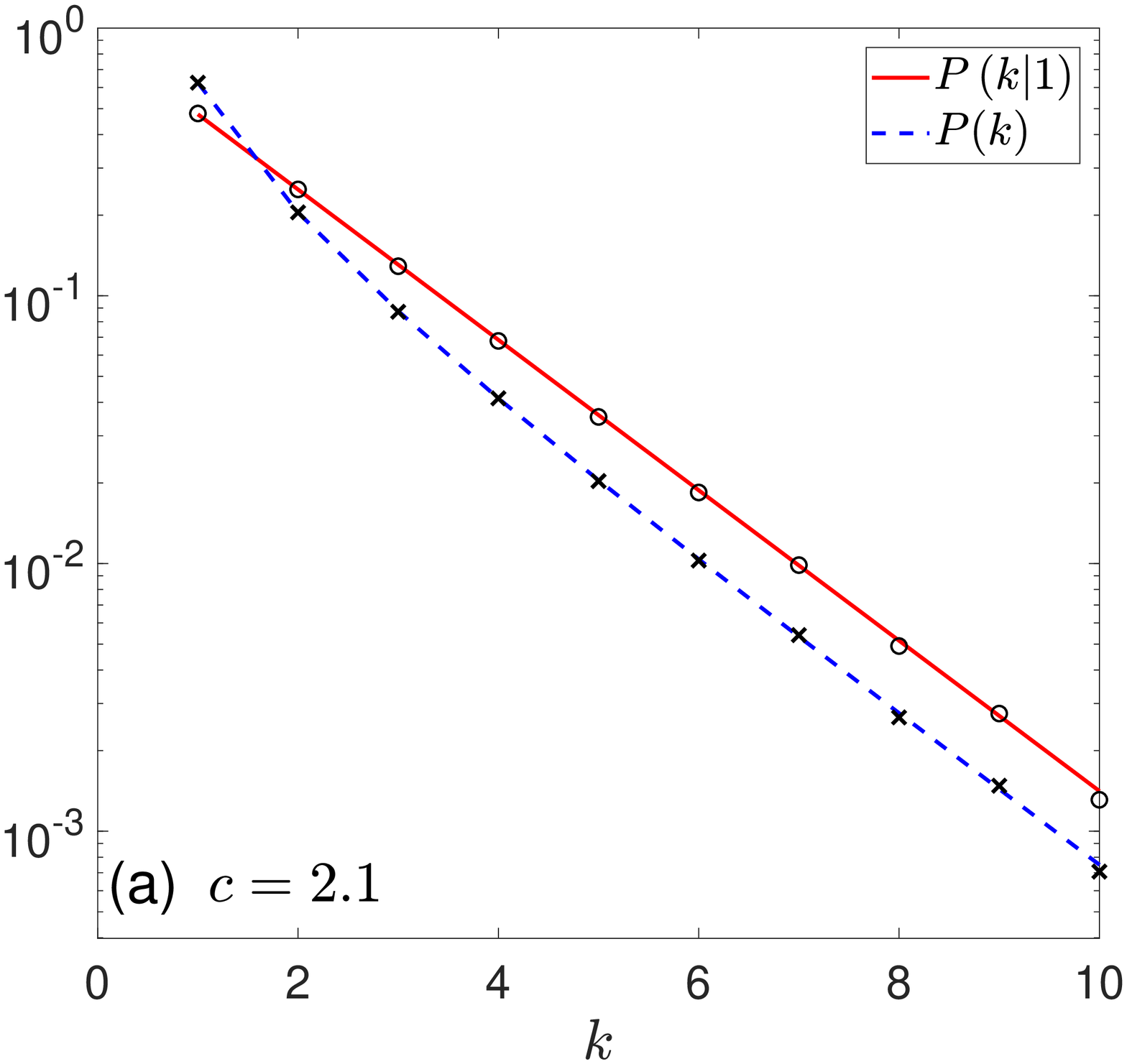} 
\hspace{0.3cm}
\includegraphics[width=6.4cm]{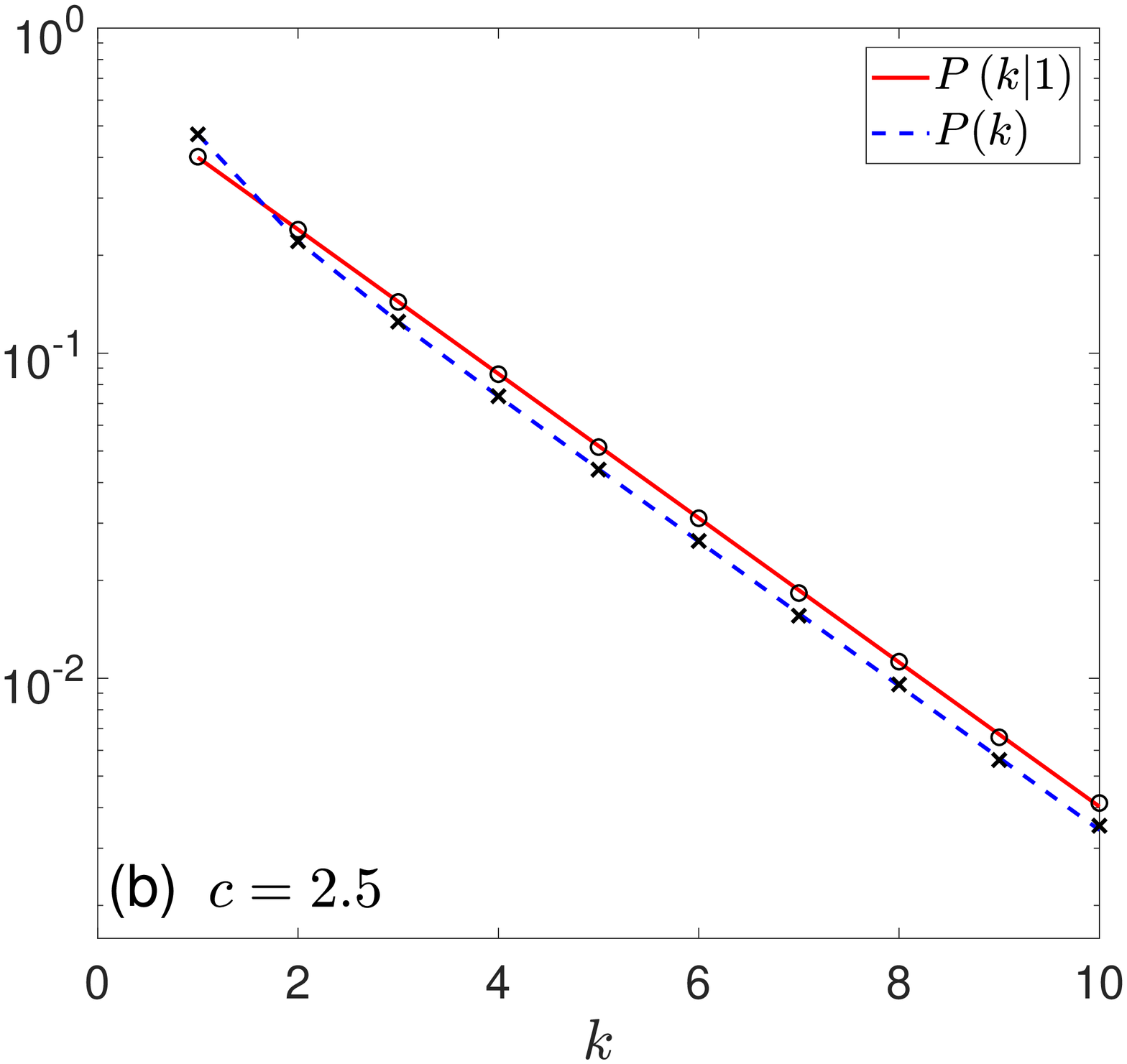} 
\\
\includegraphics[width=6.4cm]{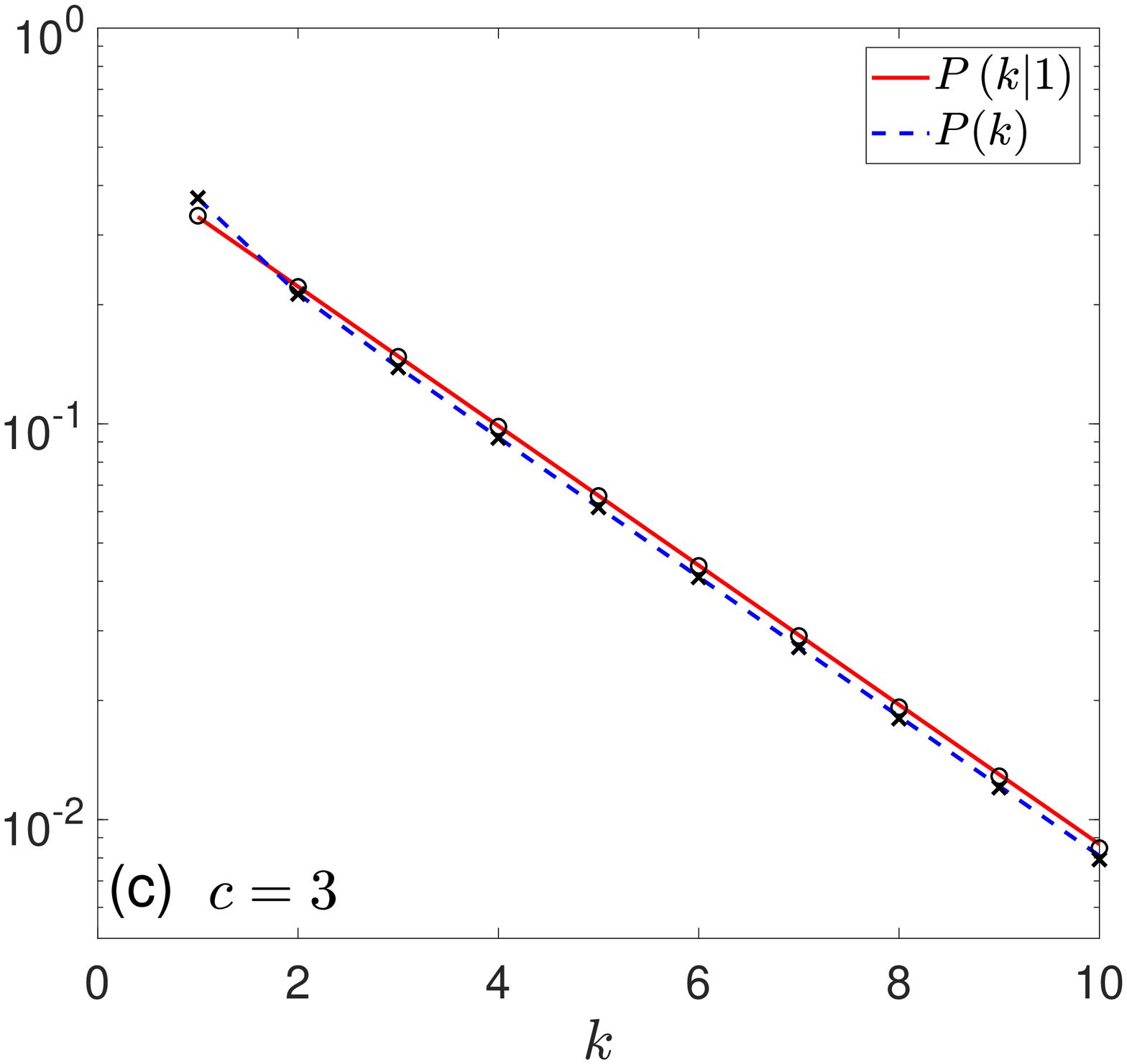} 
\end{center}
\caption{
(Color online)
Analytical results (dashed lines) for the degree distributions $P(k)$
and simulation results for the corresponding degree sequences with 
$N=10^4$ ($\times$), 
of configuration model networks 
whose giant components exhibit exponential degree distributions (solid lines) 
of the form $P(k|1)$, given by Eq. (\ref{eq:exp}), with
mean degree $c={\mathbb E}[K|1]$, where $c=2.1$ (a), $c=2.5$ (b) and $c=3.0$ (c).
The degree sequences of the resulting single-component networks (circles)
fit perfectly with the desired exponential degree distributions (solid lines).
It is found that on the giant component the abundance of nodes of degree
$k=1$ is depleted, while the abundance of nodes of higher degrees
is slightly enhanced.
This feature is most pronounced in the dilute network limit, in which
the fraction of nodes that reside on the giant components is small.
}
\label{fig:3}
\end{figure}

In Fig. \ref{fig:3} we present analytical results (dashed lines) 
for the degree distributions $P(k)$ and simulation results for 
the corresponding degree sequences ($\times$)
of the configuration model networks  
whose giant components exhibit exponential degree distributions with
mean degrees $c={\mathbb E}[K|1]$, where $c=2.1$ (a), $c=2.5$ (b) and $c=3.0$ (c). 
The degree sequences of the resulting single-component networks (circles)
fit perfectly with the desired exponential degree distributions (solid lines),
given by Eq. (\ref{eq:exp}). 
It is found that on the giant component the abundance of nodes
of degree $k=1$ is depleted with respect to their abundance in
the whole network, while the abundance of nodes of higher degrees is enhanced.

\begin{figure}
\begin{center}
\includegraphics[width=7cm]{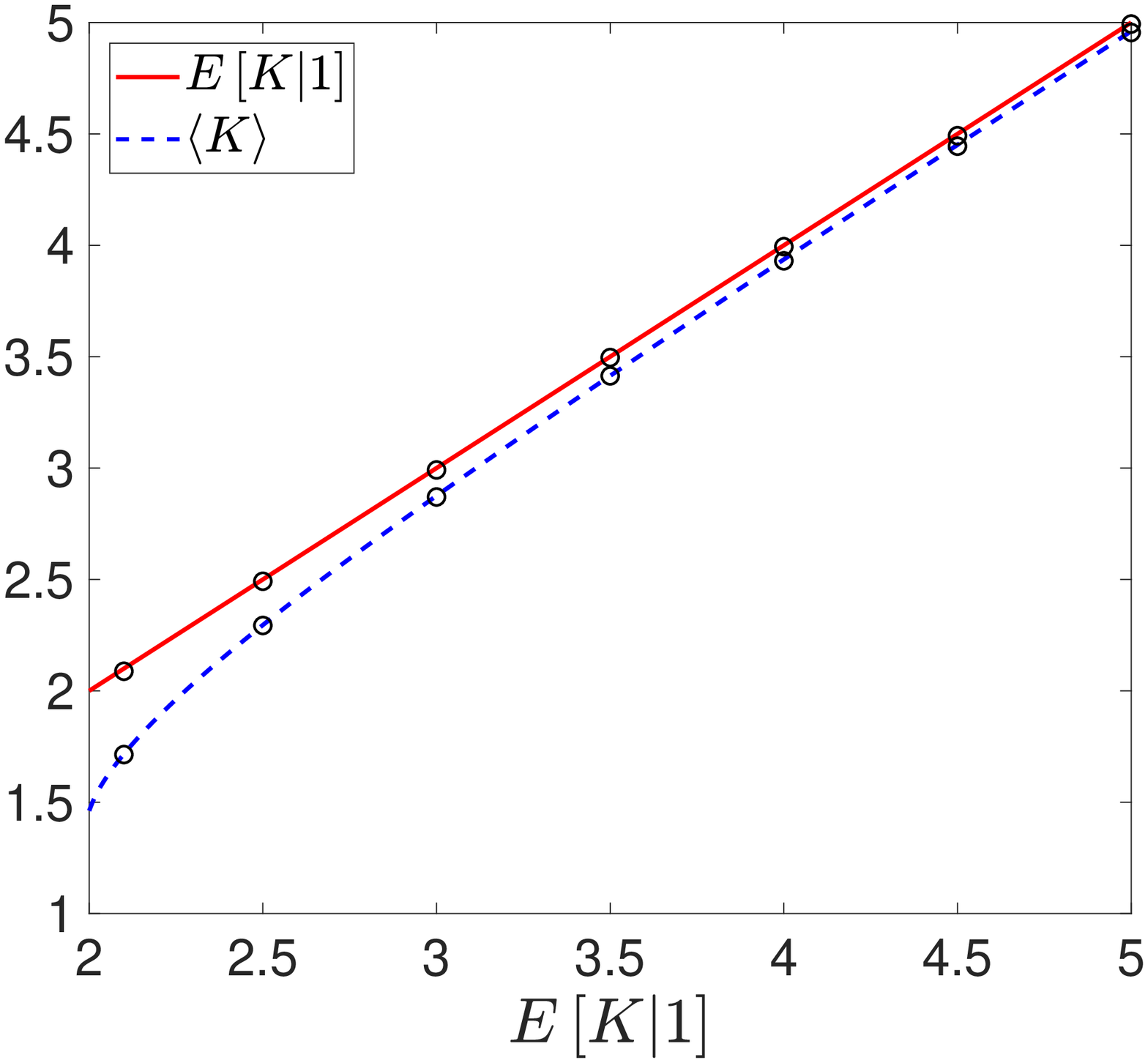} 
\end{center}
\caption{
(Color online)
Analytical results (dashed line) 
and simulation results, obtained for $N=10^4$ (circles), for
the mean degree $\langle K \rangle$ of a configuration model network whose
giant component exhibits an exponential degree distribution 
with mean degree $c={\mathbb E}[K|1]$,
as a function of ${\mathbb E}[K|1]$.
For comparison we also present the analytical results (solid line) and
simulation results (circles) for the
mean degree ${\mathbb E}[K|1]$ of the giant component.
It is found that in the dilute network limit $\langle K \rangle$
is significantly smaller than $c={\mathbb E}[K|1]$ and the two
curves converge as the network becomes denser.
}
\label{fig:4}
\end{figure}

In Fig. \ref{fig:4} we present the
mean degree $\langle K \rangle$ (dashed line),
obtained from Eq. (\ref{eq:<K>_EK}),
of a configuration model network whose
giant component exhibits an exponential degree 
distribution with mean degree $c={\mathbb E}[K|1]$,
as a function of $c$.
The mean degree, $c$, of the giant component (solid line)
is also shown for comparison.
It is found that for dilute networks $\langle K \rangle$ is significantly
smaller than $c$ and the gap between the two curves shrinks as the
network becomes denser.
The simulation results (circles),
obtained for $N=10^4$, are found to be in very good agreement
with the analytical results.

\subsection{Construction of a single component network with a power-law degree distribution}

Consider a configuration model network 
whose giant component exhibits a power-law degree distribution
of the form

\begin{equation}
P(k|1) = \frac{A}{ k^{\gamma} },
\label{eq:PLnorm1}
\end{equation}

\noindent
for $k_{\rm min} \le k \le k_{\rm max}$.
Here we focus on the case of $k_{\rm min}=1$.
In this case, the normalization coefficient is

\begin{equation}
A = \frac{1}{ \zeta(\gamma) - \zeta(\gamma,k_{\rm max}+1) },
\label{eq:PLnorm2}
\end{equation}

\noindent
where $\zeta(s,a)$ is the Hurwitz zeta function 
and $\zeta(s)=\zeta(s,1)$ is the Riemann zeta function
\cite{Olver2010}.
In order to avoid correlations, the network size must satisfy
the condition
$N > (k_{\rm max})^2/\langle K \rangle$
\cite{Bianconi2008,Bianconi2009,Janssen2015}.
The mean degree is given by

\begin{equation}
c = {\mathbb E}[K|1] = 
\frac{ \zeta(\gamma-1) - \zeta(\gamma-1,k_{\rm max}+1) }
{ \zeta(\gamma) - \zeta(\gamma,k_{\rm max}+1) }.
\label{eq:Kmsf}
\end{equation}

\noindent
As noted above, a single connected component with a degree distribution
$P(k|1)$ exists only if the condition $\mathbb{E}[K|1] \ge 2$ is satisfied.
This implies that for a given value of $k_{\rm max}$ there exists a critical
value of $\gamma$, denoted by $\gamma_c(k_{\rm max})$, such that a 
giant component exists only for $\gamma < \gamma_c(k_{\rm max})$.
The value of $\gamma_c(k_{\rm max})$ is obtained by solving 
Eq. (\ref{eq:Kmsf}) for $\gamma$ under the condition that $c=2$.
In the special case of $k_{\rm max} \rightarrow \infty$
one obtains
$\gamma_c(k_{\rm max}) \rightarrow \gamma_c(\infty) = 3.4787...$,
which is a solution of the equation 
$\zeta(\gamma-1) = 2 \zeta(\gamma)$.

The second moment of the degree distribution is

\begin{equation}
{\mathbb E}[K^2|1] = 
\frac{ \zeta(\gamma-2) - \zeta(\gamma-2,k_{\rm max}+1) }
{ \zeta(\gamma) - \zeta(\gamma,k_{\rm max}+1) }.
\label{eq:K2msf}
\end{equation}

\noindent
For $\gamma \le 2$,
in the asymptotic limit of $N \rightarrow \infty$,
the mean degree ${\mathbb E}[K|1]$ diverges 
in the limit
$k_{\rm max} \rightarrow \infty$.
For $2 < \gamma \le 3$, in the asymptotic limit,
the mean degree is 
bounded while the second moment
${\mathbb E}[K^2|1]$ diverges.
For $\gamma > 3$ both moments are bounded.
The generating functions of $P(k|1)$
for a giant component with a power-law degree distribution are

\begin{equation}
G_0^1(x) = 
\frac{ {\rm Li}_{\gamma}(x) - x^{k_{\rm max}+1} 
\Phi(x,\gamma,k_{\rm max}+1) }{\zeta(\gamma) - \zeta(\gamma,k_{\rm max}+1)}
\label{eq:G0sf}
\end{equation}

\noindent
and

\begin{equation}
G_1^1(x) = 
\frac{ {\rm Li}_{\gamma-1}(x)  -  x^{k_{\rm max}+1} 
\Phi(x,\gamma-1,k_{\rm max}+1)}
{x \left[\zeta(\gamma-1) - \zeta(\gamma-1,k_{\rm max}+1) \right]},
\label{eq:G1sf}
\end{equation}

\noindent
where ${\rm Li}_{\gamma}(x)$
is the polylogarithmic function.
Inserting the expressions for the two 
generating functions into Eq. (\ref{eq:gtimplicit}),
we obtain

\begin{eqnarray}
\tilde g (2-\tilde g) 
\left[ 1 + (1-\tilde g)
\frac{ {\rm Li}_{\gamma-1}(1-\tilde g)  -  (1-\tilde g)^{k_{\rm max}+1} 
\Phi(1-\tilde g,\gamma-1,k_{\rm max}+1)}
{(1-\tilde g) \left[\zeta(\gamma-1) - \zeta(\gamma-1,k_{\rm max}+1) \right]}
\right.
\nonumber \\
\left.
- \frac{{\rm Li}_{\gamma}[(1-\tilde g)^{3/2} - (1-\tilde g)^{3(k_{\rm max}+1)/2} 
\Phi[(1-\tilde g)^{3/2},\gamma,k_{\rm max}+1]   } {\ln (1-\tilde g)  
 [\zeta(\gamma-1) - \zeta(\gamma-1,k_{\rm max}+1)] }
\right]
=1.
\label{eq:gtpowerlaw}
\end{eqnarray}

\noindent
This is an implicit equation for $\tilde g$ in terms of the exponent $\gamma$
and the upper cutoff $k_{\rm max}$, that should be solved numerically.
The parameter $g$ is then obtained from Eq. (\ref{eq:gshort}).
Inserting $G_0^1(x)$ from Eq. (\ref{eq:G0sf}) into Eq. (\ref{eq:gshort}),
we obtain

\begin{equation}
g = \left[ 1 + \sum\limits_{n=1}^{\infty} 
\frac{ {\rm Li}_{\gamma}[(1-\tilde g)^n] - (1-\tilde g)^{n(k_{\rm max}+1)} 
\Phi[(1-\tilde g)^n,\gamma,k_{\rm max}+1] }{\zeta(\gamma) - \zeta(\gamma,k_{\rm max}+1)}
\right]^{-1}.
\label{eq:gpowerlaw}
\end{equation}

In order to generate an ensemble 
of single component networks 
whose mean size is $\langle N_1 \rangle$, 
which exhibit
a given power-law degree 
distribution $P(k|1)$,
one generates configuration model networks
of size $N = \langle N_1 \rangle/g$
with the degree
distribution $P(k)$, 
given by Eq. (\ref{eq:pk2}),
where $\tilde g$ is given by Eq. (\ref{eq:gtpowerlaw}),
$g$ is given by Eq. (\ref{eq:gpowerlaw})
and $P(k|1)$ is given by Eq. (\ref{eq:PLnorm1}).
Note that for $\gamma \ge 2$, in the limit of $k_{\rm max} \rightarrow \infty$
one obtains that $g \rightarrow g_{\infty} < 1$. 
This means that in configuration model networks which exhibit a power-law degree
distribution with $\gamma \ge 2$ the giant component does not encompass the whole
network regardless of the value of $k_{\rm max}$.
This means that the approach presented here is applicable and useful for the construction
of single component random networks with power-law degree distributions
for the whole range of $2 \le \gamma \le \gamma_c(\infty)$.

\begin{figure}
\begin{center}
\includegraphics[width=7cm]{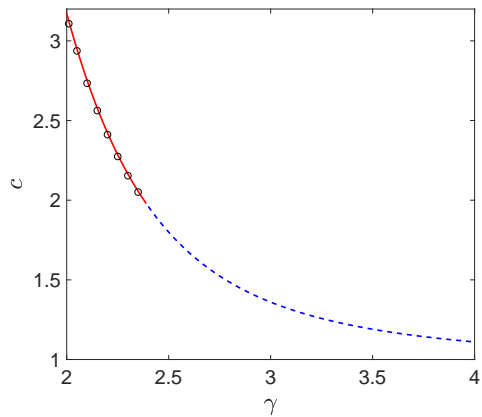} 
\end{center}
\caption{
(Color online)
The mean degree, $c={\mathbb E}[K|1]$, 
of the giant component of a configuration model network
(solid line) with a power-law degree distribution
[Eq. (\ref{eq:PLnorm1})],
as a function of the exponent $\gamma$,
for $\gamma \ge 2$ with $k_{\rm max}=100$,
given by Eq. (\ref{eq:Kmsf}). 
The mean degree decreases as $\gamma$ is increased.
For $\gamma > 2.4$ the solid line is replaced by a dashed line, 
which is still a solution of Eq. (\ref{eq:Kmsf}). 
However, it does not describe the mean degree 
of a giant component, because in this regime $c < 2$ while
the mean degree of a giant component must satisfy $c \ge 2$.
The results for the mean degrees of the single component networks constructed
using this method (circles) are in perfect agreement with the analytical results.
}
\label{fig:5}
\end{figure}

In Fig. \ref{fig:5} we present analytical results (solid line),
obtained from Eq. (\ref{eq:Kmsf}),
for the mean degree, 
$c={\mathbb E}[K|1]$, of the giant component of a configuration
model network, for which the giant component exhibits a power-law degree
distribution, $P(k|1)$, given by Eq. (\ref{eq:PLnorm1}),
as a function of the exponent $\gamma$ for $2 < \gamma < 2.4$.
The upper cutoff of the degree distribution is $k_{\rm max}=100$.
The dashed line, presented for $\gamma > 2.4$, is still a solution of 
Eq. (\ref{eq:Kmsf}). However, it does not describe the mean degree 
of a giant component, because in this regime $c < 2$ while
the degree distribution of a giant component must satisfy $c>2$.
The results for the mean degrees of the network instances constructed
using this method (circles) are in perfect agreement with the analytical results.
It is found that the mean degree decreases as $\gamma$ is increased.

\begin{figure}
\begin{center}
\includegraphics[width=7cm]{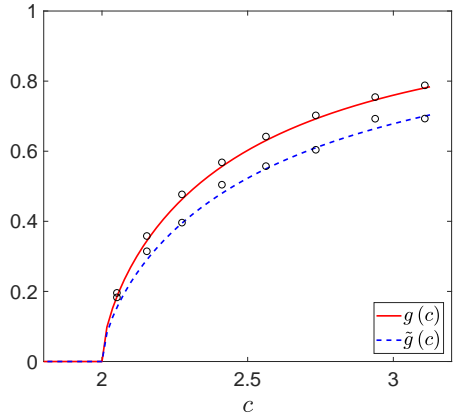} 
\end{center}
\caption{
(Color online)
The parameters $g$ (solid line) and $\tilde g$ (dashed line) of a configuration model network
whose giant component exhibits a power-law degree distribution
of the form $P(k|1)$, given by Eq. (\ref{eq:PLnorm1}),
as a function of the mean degree $c={\mathbb E}[K|1]$ of the giant component.
As discussed in the text the minimal value of the mean degree of a 
giant component with a power-law degree distribution is $c=2$.
Thus, for $c<2$ both $g=0$ and $\tilde g=0$. For $c>2$ 
the parameters
$g$ and $\tilde g$ gradually increase.
This is in contrast to the case of the exponential degree distribution,
shown in Fig. \ref{fig:2}, in which $g$ and $\tilde g$ increase more steeply.
}
\label{fig:6}
\end{figure}

In Fig. \ref{fig:6} we show analytical results for the values of
the parameters $g$ (solid line) and $\tilde g$ (dashed line) of a configuration model network
whose giant component exhibits a power-law degree distribution,
as a function of the mean degree $c={\mathbb E}[K|1]$ of the giant component.
As discussed above, both $g$ and $\tilde g$ vanish for $c<2$, 
since there are no giant components with mean degrees
lower than $2$. 
For $c>2$ the parameters
$g$ and $\tilde g$ gradually increase.
This is in contrast to the case of the exponential degree distribution,
shown in Fig. \ref{fig:2}, in which $g$ and $\tilde g$ increase more steeply.
The simulation results (circles) for $g$, obtained from network instances constructed using
this method with
$k_{\rm max}=100$ 
and
$N=4 \times 10^4$ 
are found to be in 
good agreement with the analytical results, while the results for
$\tilde g$ are a bit noisy.

\begin{figure}
\begin{center}
\includegraphics[width=6.4cm]{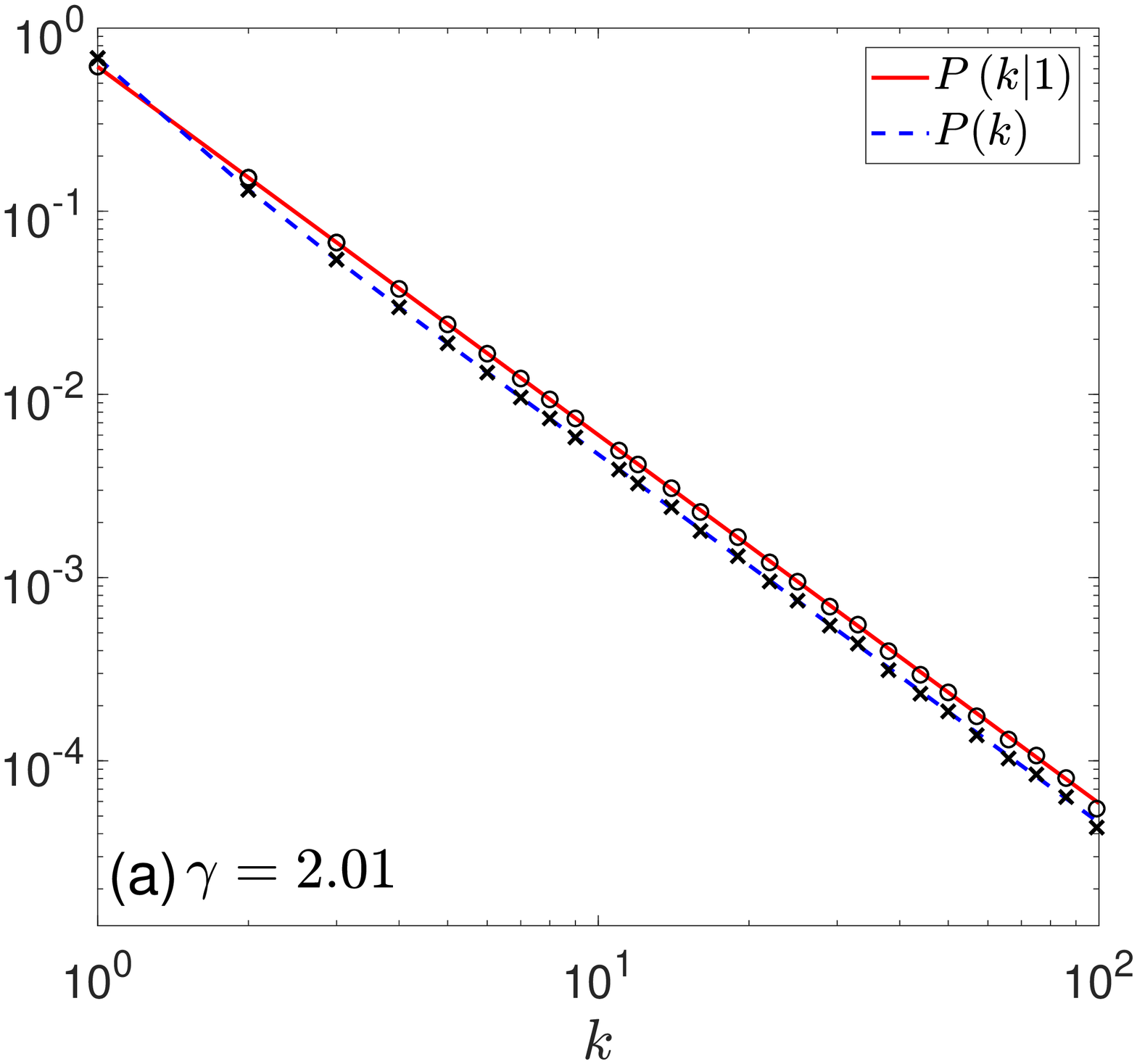} 
\hspace{0.3cm}
\includegraphics[width=6.4cm]{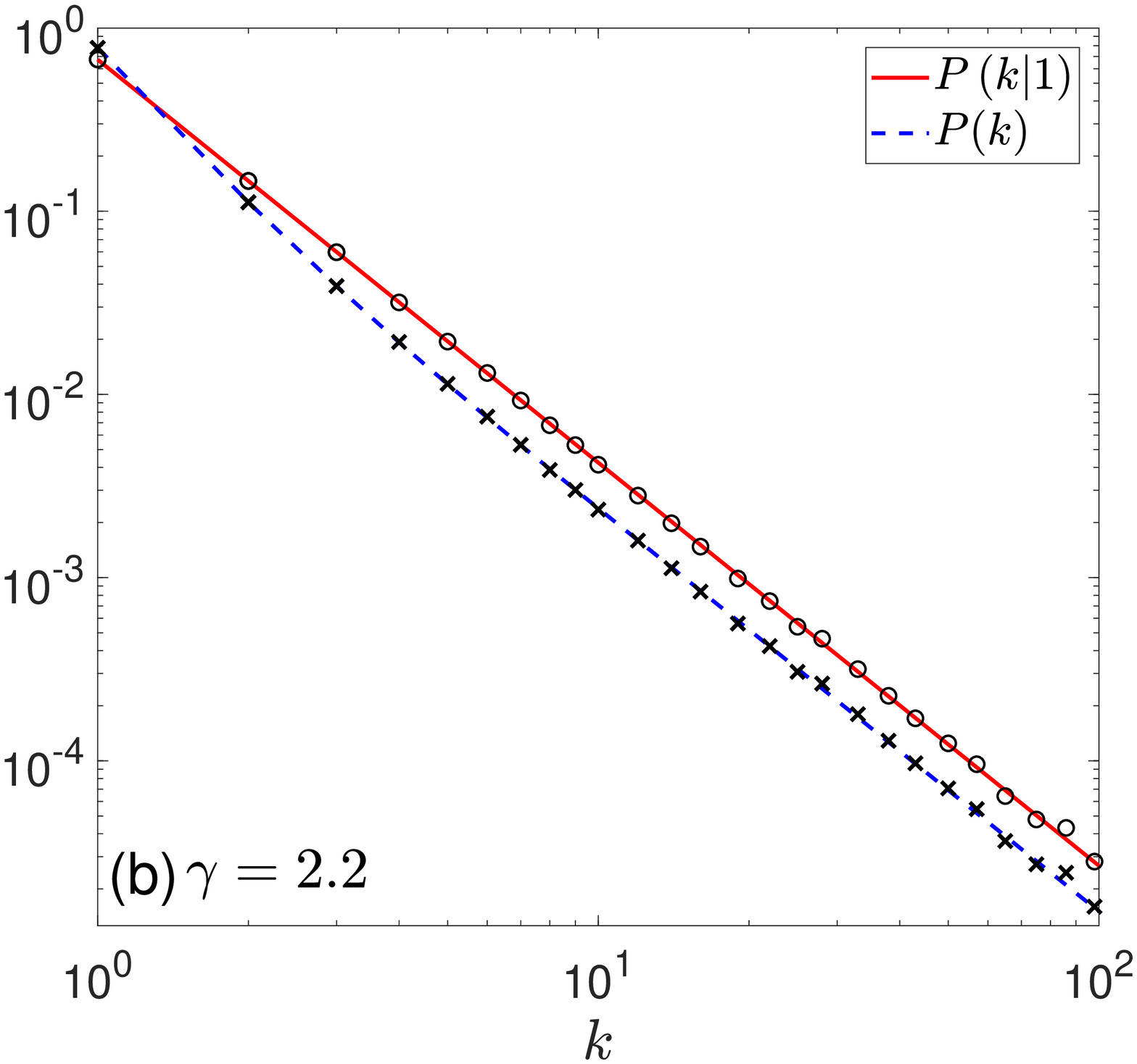}  
\\
\includegraphics[width=6.4cm]{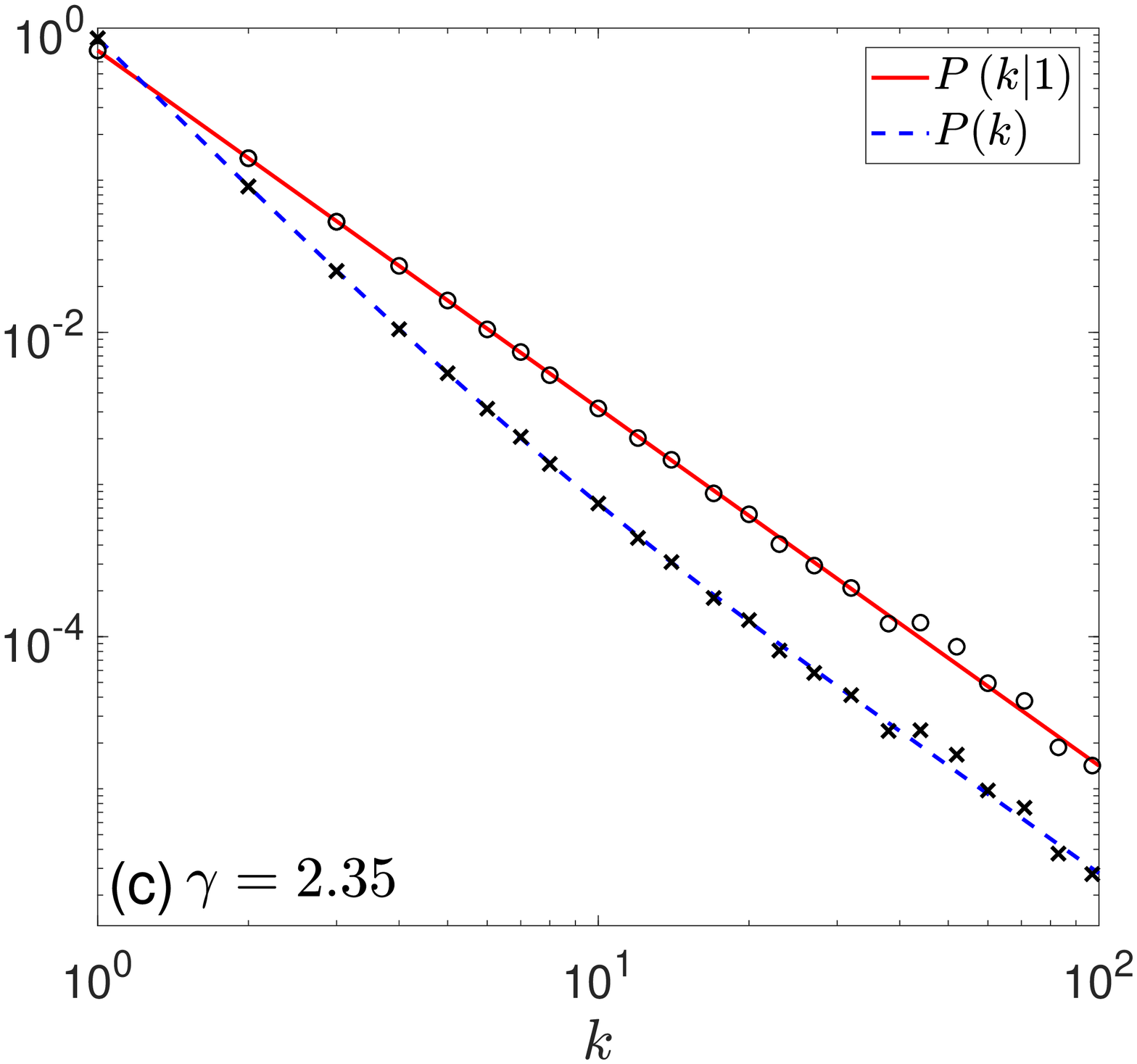} 
\end{center}
\caption{
(Color online)
Analytical results (dashed lines) for the degree distributions $P(k)$ 
and simulation results with $N=4 \times 10^4$ for the corresponding
degree sequences ($\times$)
of configuration model networks
whose giant components exhibit power-law degree distributions (solid lines),
of the form $P(k|1)$, given by Eq. (\ref{eq:PLnorm1}), with
$\gamma=2.01$ (a), $\gamma=2.2$ (b) and $\gamma=2.35$ (c), 
and with $k_{\rm max}=100$.
The degree sequences of the resulting single-component networks (circles),
fit perfectly with the desired power-law degree distributions (solid lines).
It is found that on the giant component the abundance of nodes of degree
$k=1$ is depleted, while the abundance of nodes of higher degrees
is enhanced.
This feature is most pronounced in the dilute network limit, in which
the fraction of nodes that reside on the giant components is small.
}
\label{fig:7}
\end{figure}

In Fig. \ref{fig:7} we present analytical results (dashed lines)
 for the degree distributions $P(k)$ 
[given by Eq. (\ref{eq:pk2}),
where $\tilde g$ is the solution of Eq. (\ref{eq:gtpowerlaw}) and
$g$ is given by Eq. (\ref{eq:gpowerlaw})]
and simulation results for the corresponding degree sequences
($\times$) of the configuration model networks 
whose giant components exhibit power-law degree distributions,
with $\gamma=2.01$ (a), $\gamma=2.2$ (b) and $\gamma=2.35$ (c). 
The degree sequences of the resulting single-component networks (circles)
fit perfectly with the desired power-law distributions (solid lines),
given by Eq. (\ref{eq:PLnorm1}).

\begin{figure}
\begin{center}
\includegraphics[width=7cm]{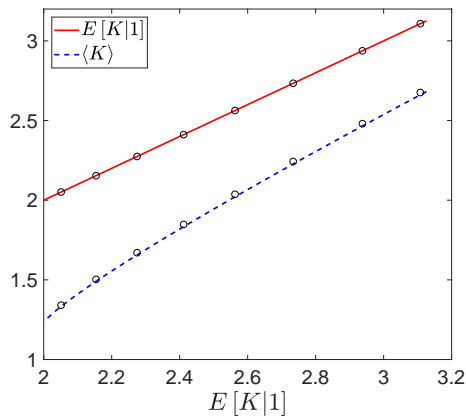} 
\end{center}
\caption{
(Color online)
The mean degree $\langle K \rangle$ of a configuration model network whose
giant component exhibits a power-law degree distribution 
with mean degree $c={\mathbb E}[K|1]$,
as a function of ${\mathbb E}[K|1]$ (dashed line).
The mean degree ${\mathbb E}[K|1]$ of the giant component
(solid line), is also shown for comparison.
It is found that in the dilute network limit $\langle K \rangle$
is much smaller than ${\mathbb E}[K|1]$.
The gap between the two curves slightly decreases as the
network becomes more dense, but the two curves do not converge.
The simulation results (circles),
obtained for $N=4 \times 10^4$,
are in very good agreement with the
analytical results. 
}
\label{fig:8}
\end{figure}

In Fig. \ref{fig:8} we present analytical results (dashed line) for the
mean degree $\langle K \rangle$ of a configuration model network whose
giant component exhibits a power-law degree distribution,
given by Eq. (\ref{eq:PLnorm1}) with $k_{\rm max}=100$, 
as a function of the mean degree 
$c={\mathbb E}[K|1]$
of the giant component. 
The mean degree $c$ of the giant component
(solid line), is also shown for comparison.
It is found that in the dilute network limit $\langle K \rangle$
is much smaller than $c={\mathbb E}[K|1]$.
The gap between the two curves slightly decreases as the
network becomes more dense, but the two curves do not converge. 
This is due to the fact that even for the largest value of 
${\mathbb E}[K|1]$ that can be obtained with
$k_{\rm max}=100$ the giant component does not encompass the
whole network. The gap between $\langle K \rangle$ can be decreased
further by increasing the value of $k_{\rm max}$.
However, in order to maintain the whole network uncorrelated its size $N$
should satisfy $N > (k_{\rm max})^2/\langle K \rangle$ 
\cite{Bianconi2008,Bianconi2009,Janssen2015}.
The results obtained from computer simulations (circles)
with $N=4 \times 10^4$ are found to be 
in very good agreement with the analytical results.

\section{Discussion}
 
While configuration model networks are random and uncorrelated,
their giant components exhibit correlations between the degrees
of adjacent nodes.
These degree-degree correlations 
and the assortativity coefficients
of the giant components were studied in Ref. 
\cite{Tishby2018}.
The giant components were found to be disassortative, namely high-degree nodes
tend to connect preferentially to low-degree nodes
and vice versa.
Moreover, it was found that as the network approaches the percolation transition
from above and the giant component decreases in size, its structure becomes more
distinct from the structure of the overall network.
In particular, the degree distribution of the giant component deviates more strongly from the
degree distribution of the whole network, 
the degree-degree correlations become stronger
and the assortativity coefficient becomes more negative.

The disassortativity of the giant component helps to maintain its integrity.
For example, the probability of a pair of nodes of degrees $k,k'=1$, which reside on the giant
component, to connect to each other must vanish, otherwise they will form an isolated dimer.
This means that nodes of degree $k=1$ preferentially connect to nodes of higher degrees.
As a result, high-degree nodes preferentially connect to nodes of degree $k=1$.
In fact, the giant component exhibits degree-degree correlations of all orders.
These correlations are required in order to exclude the possibility that a randomly
selected node belongs to an isolated component of any finite size
\cite{Tishby2018}.
Interestingly, disassortativity was found to be prevalent in a broader class of
scale-free networks which exhibit correlations and can be explained by
entropic considerations
\cite{Johnson2010,Williams2014}. 

The methodology introduced in this paper enables the construction of 
random networks that consist of a single connected component of $N_1$
nodes with a given degree distribution $P(k|1)$.
The desired network consists of the giant component of a suitable
configuration model network of $N$ nodes and degree distribution $P(k)$.
For a given value of $N$ the size $N_1$ of the giant component exhibits
fluctuations which satisfy ${\rm Var}(N_1) \propto N$, which are thus
under control in the asymptotic limit. 
We also present an adjustment procedure for
the case in which a specific value of $N_1$ is required.

The construction of random networks that consist of a single connected component
with a given degree distribution is expected to be useful for the
analysis of empirical networks.
A common practice in the study of empirical networks is to generate
an ensemble of randomized networks with the same degree sequence
as the empirical network. One then compares structural and statistical
properties of the empirical network to the corresponding properties
of the randomized networks. The differences between the empirical
network and its randomized counterparts may imply some 
significant functional or evolutionary properties of the empirical
network. 
Stated more technically, randomized networks serve as null 
models for empirical networks 
\cite{Bianconi2008,Bianconi2009,Coolen2009,Annibale2009,Roberts2011,Roberts2013,Coolen2017}.
This approach was utilized in the study of network
motifs, which are over-represented in empirical networks
compared to the corresponding randomized networks
\cite{Shen2002,Kashtan2004}.
It was also used in the analysis of degree-degree correlations,
the assortativity coefficient and the clustering coefficient
\cite{Maslov2004,Park2003,Holme2007},
and in the study of the distribution of shortest path lengths
\cite{Giot2003}.

A randomized network with the same degree sequence as a
given empirical network can be constructed in two different ways.
One way is to generate a configuration model network with the
given degree sequence obtained from the empirical network. 
Another way is to start from the empirical network and apply
a series of rewiring steps. In each rewiring step one picks two
random edges, $i-j$ and $i'-j'$ and then exchanges them
such that $i$ becomes connected to $j'$ and $i'$ becomes 
connected to $j$. In a case in which either the $i-j'$ edge or
the $i'-j$ edge already exists the step is rejected.
After a large number of such rewiring steps one obtains
a randomized network which maintains the degree 
sequence of the empirical network.

In some cases one may be interested in finding the 
degree distribution from which the given degree
sequence of the empirical network is most likely
to arise. Consider an empirical network of $N$
nodes, whose degree sequence is given by
$\{ n_k^{\rm E} \}$, $k=1,2,\dots,k_{\rm max}$,
where $n_k^{\rm E}$ is the number of nodes of degree $k$
and $\sum_k n_k^{\rm E} = N$.
The degree distribution from which this degree sequence
is most likely to emerge is given by

\begin{equation}
P(k) = \frac{n_k^{\rm E}}{N}, 
\label{eq:Pkemp}
\end{equation}

\noindent
where
$k=1,2,\dots,k_{\rm max}$.
Sampling the degrees of $N$ nodes from this distribution,
the probability to obtain a degree sequence of the
form $\{ n_k \}$, $k=1,2,\dots,k_{\rm max}$ is

\begin{equation}
P(\{n_k\}) = \frac{N!}{\prod_{k=1}^{k_{\rm max}} n_k!}
\prod_{k=1}^{k_{\rm max}} P(k)^{n_k}.
\end{equation}

\noindent
Configuration model networks with degree sequences
that are drawn from the degree distribution $P(k)$, given by Eq. 
(\ref{eq:Pkemp}), provide a broader class of randomized networks
for the given empirical networks. While their degree sequences
are not identical to the degree sequence of the empirical network
their statistical properties are closely related. 
This is a grand-canonical approach to the sampling problem.

While some empirical networks consist of a single connected component such as transportation networks and brain networks \cite{Wandelt2019},
other networks consist of many isolated components of various sizes such as adoption of innovations or products networks \cite{Karsai2016} and  mobile phone calling networks \cite{Li2014}. 
The distribution of sizes of these components has been studied in the context of
subcritical networks and provides a useful characterization of the network
structure 
\cite{Katzav2018}. 
In a case in which one of the isolated components is particularly large
(and asymptotically encompasses a macroscopic fraction of the network
size), it is referred to as the giant component. In such case the network
exhibits a coexistence between the giant component and many finite components.
Here we focus on the properties of the giant component,
namely the degree distribution, degree-degree correlations, clustering  
coefficient and size. The size of the giant component, $N_1$, depends on the
size of the whole network, $N$, and on the fraction of nodes, $0 < g < 1$,
that reside on the giant component. In computer simulations the value of
$g$ varies between different network instances in a given network ensemble,
following a distribution $P(g)$ that is characteristic of the given ensemble.
In empirical networks it is difficult to find many network instances that are
drawn from the same statistical ensemble. Therefore, it is difficult to find
a direct analog of $P(g)$ in empirical networks.

In a case in which the empirical network under study consists of
a single connected component, it is desirable that the
corresponding randomized networks will also consist of
a single connected component. The procedures described above may
produce randomized networks that consist of multiple components
(such as a giant component and many finite components),
even in a case in which the empirical network consists of a single 
connected component. The size of the giant component 
of the randomized network depends on its degree sequence 
and can be determined using methods of percolation theory.

The methodology presented in this paper provides a way to
obtain a randomized network that consists of a single connected
component. Consider an empirical network of $N_1$ nodes that
consists of a single connected component with degree sequence
$\{ n_k \}$. Using Eq. (\ref{eq:Pkemp}) one obtains the most
probable degree distribution $P(k|1)$ for the given degree 
sequence. Using the procedure presented in this paper, one
obtains the size $N$ and the degree distribution $P(k)$ of
a configuration model network whose giant component 
is the desired randomized network.

\section{Summary}

We presented a method for the construction of ensembles of random networks 
that consist of a single connected component of any desired size $N_1$ 
with a pre-defined degree distribution $P(k|1)$.
The construction is done by generating a configuration model network
with a suitable degree distribution $P(k)$ and size $N$,
whose giant component is of size $N_1$ and its degree distribution is $P(k|1)$.
This approach is based on the inversion of the relation between $P(k)$ and $P(k|1)$,
which was presented in Ref. 
\cite{Tishby2018}.
It extends the construction toolbox of random networks beyond the
configuration model framework, in which one controls the network size and the 
degree distribution but has no control over
the number of network components and their sizes.
The capability of generating single component random networks with a
desired degree distribution is expected to be instrumental in the effort
to elucidate the statistical properties of such networks at the local and
global scales.

\end{document}